\documentstyle[12pt,aaspp4,tighten]{article}

\def\etal{{\it et al.~}}
\def\k{{\kappa}}

\def\VEV#1{\left\langle #1\right\rangle}
\def\abso#1{\mid\! #1\!\mid}
\def\la{\hbox{ \raise.35ex\rlap{$<$}\lower.6ex\hbox{$\sim$}\ }}
\def\ga{\hbox{ \raise.35ex\rlap{$>$}\lower.6ex\hbox{$\sim$}\ }}
\def\bftheta{{\mbox{\boldmath $\theta$}}}
\def\bfkappa{{\mbox{\boldmath $\kappa$}}}
\def\bfw{{\mbox{\boldmath $w$}}}
\def\bfx{{\mbox{\boldmath $x$}}}
\def\bfr{{\mbox{\boldmath $r$}}}
\def\bfk{{\mbox{\boldmath $k$}}}
\def\W2{{\cal W}}

\def\veck{{\mbox{\boldmath $k$}}}
\def\hattheta{{\bf \hat \theta}}

\slugcomment{CAL-682, CU-TP-936, astro-ph/9903486}

\begin{document}

\title{The Angular Three-Point Correlation
Function in the Quasilinear Regime}

\author{Ari Buchalter$^*$\altaffilmark{1}, Marc
Kamionkowski$^\dagger$\altaffilmark{2}, and Andrew H. Jaffe
$^\ddagger$\altaffilmark{3}}
\affil{$^*$Department of Astronomy and Columbia Astrophysics
     Laboratory, Columbia University, 550 West 120th St., New
     York, NY~~10027}
\affil{$^\dagger$Department of Physics and Columbia Astrophysics 
     Laboratory, Columbia University, 550 West 120th St., New
     York, NY~~10027}
\affil{$^\ddagger$Center for Particle Astrophysics, 301 LeConte
Hall, University of California, Berkeley, CA~~94720}
\altaffiltext{1}{ari@astro.columbia.edu}
\altaffiltext{2}{kamion@phys.columbia.edu}
\altaffiltext{3}{jaffe@cfpa.berkeley.edu}

\begin{abstract}
We calculate the normalized angular three-point correlation 
function (3PCF), $q$, as well as the normalized angular skewness, $s_3$, 
assuming the small-angle approximation, for a
biased mass distribution in flat and open cold-dark-matter (CDM) models 
with Gaussian initial conditions. The leading-order 
perturbative results incorporate the explicit dependence on the
cosmological parameters, 
the shape of the CDM transfer function, the linear evolution
of the power spectrum, the form of redshift distribution function,
and linear and nonlinear biasing, which may be evolving. 
Results are presented for
different redshift distributions, including that appropriate for
the APM Galaxy Survey, as well as for a survey with a mean redshift
of $\overline{z} \simeq 1$ (such as the VLA FIRST Survey).
Qualitatively, many of the results found for $s_3$ and $q$
are similar to those obtained in a related treatment of the spatial 
skewness and 3PCF (Buchalter \& Kamionkowski 1999), such as
a leading-order correction to the
standard result for $s_3$ in the case of nonlinear bias (as defined
for unsmoothed density fields),
and the sensitivity of the configuration dependence of $q$
to both cosmological and biasing models.
We show that since angular CFs are 
sensitive to clustering over a range of redshifts, 
the various evolutionary dependences
included in our predictions 
imply that measurements
of $q$ in a deep survey might better discriminate
between models with different histories, such as 
evolving vs. non-evolving bias, that can have similar spatial
CFs at low redshift. 
Our calculations employ a derived equation---valid for open, closed,
and flat models---for obtaining the angular bispectrum from the spatial
bispectrum in the small-angle approximation.

\end{abstract}

\keywords{large scale structure of the universe --- cosmology: theory --- 
galaxies: clustering --- galaxies: statistics}

\section{INTRODUCTION}

Characterization of the initial distribution of density perturbations
and understanding their subsequent evolution into present-day structures
are among the central aims of cosmology. 
Particular attention has been focused on the low-redshift
distribution of matter over large scales, 
which can probe the linear physics of the early
universe and therefore 
test models for the origin of large-scale structure, 
such as inflation.
The $n$-point correlation functions\footnote{The 
term ``correlation function'' is understood throughout to refer
to the connected, or reduced, correlation function.
Note that the $n$-point CFs are valid statistics only if the universe
is homogeneous on large scales.
We invoke
the assumptions of large-scale homogeneity and isotropy, 
so that the CFs depend
only relative distances and are independent of orientation.} (CFs)
have become the most widely used statistical
tools to quantify the observed distribution of matter, 
in part because they can be easily related
to the predictions of such models and thus discriminate between them.
In principle, recovering the current distribution of density
perturbations is straightforward; one simply
maps the three-dimensional distribution of mass with a
sufficiently large redshift survey.  
Existing redshift surveys, however, typically contain relatively small numbers
of objects, have inconvenient shapes, and necessarily suffer
from redshift-space distortions, 
all of which make it difficult to obtain precise CF measurements, particularly
for the higher-order functions. 

Another possibility is to map the angular distribution of galaxies as 
projected onto the surface of the sky.
Angular surveys 
can sample larger volumes of space and contain greater numbers of objects.
Furthermore, the effects of redshift distortions
need not be considered in an angular survey.
Thus, if the redshift distribution is well constrained, as is often
the case for large angular surveys, angular CFs can in practice
provide far greater statistical power and more insight into the evolution
of clustering, than sparser, redshift-survey data 
(Groth \& Peebles 1977; Fry \& Seldner 1982; 
Jing, Mo, \& B\"{o}rner 1991; Baugh \& Efstathiou 1993; 
Cappi \& Maurogordato 1995; 
Cress \etal 1996; Maddox, Efstathiou \& Sutherland 1996;
Gazta\~{n}aga \& Baugh 1998; 
Cress \& Kamionkowski 1998).

To date, angular-clustering studies
have focussed primarily on the 
angular two-point correlation function (2PCF), $\varpi(\theta)$, but
with advent of new, deep 
surveys at various wavelengths (radio, optical, IR, x-ray)
and more powerful computing resources, increasing attention is
being focused on higher-order CFs.
These are particularly important in assessing primordial Gaussianity
(Fry 1984; Goroff \etal 1986; Bernardeau 1994;
Fry \& Scherrer 1994; Gazta\~{n}aga \& Bernardeau 1998; White 1998;
Scoccimarro \etal 1998; Scoccimarro, Couchman, \& Frieman 1998), 
since to leading order they, unlike the 2PCF,
reflect the inherently nonlinear gravitational amplification of the
initial fluctuations.
Furthermore, they contain 
information about the cosmological parameters (Bouchet \etal 1992; 
Juszkiewicz, Bouchet, \& Colombi 1993; Gazta\~{n}aga, 
Croft, \& Dalton 1995; Bouchet \etal 1995; Catelan \etal 1995; Martel 1995;
Jing \& B\"{o}rner 1997; Szapudi \etal 1998; Kamionkowski \& Buchalter 1998) 
and about the nonlinear bias between the observed and underlying
density distributions (Jensen \& Szalay 1986; Szalay 1988;
Fry \& Gazta\~{n}aga 1993; Frieman \& Gazta\~{n}aga 1994; Fry 1994;
Jing 1997; Matarrese, Verde, \& Heavens 1997), 
which cannot be obtained from $\varpi(\theta)$.

In particular, the angular three-point correlation
function (3PCF), $Z(\theta_{12},\theta_{23},\theta_{31})$,
and relatedly, the normalized angular skewness, $s_3(\Theta)$
(proportional to $Z$ evaluated at zero lag, for a distribution
smoothed over angular scale $\Theta$),
are the lowest-order intrinsically nonlinear statistics and
can therefore yield important constraints on models of structure formation. 
Moreover, if different populations are differently biased relative
to the underlying mass distribution
then measuring the angular 3PCFs of these
populations can provide multiple, complementary constraints. 
Measurements
of $s_3$, as well as the normalized angular 3PCF, 
$q({\theta}_{12},{\theta}_{23},{\theta}_{31}) = 
Z({\theta}_{12},{\theta}_{23},{\theta}_{31})/\left[
\varpi(\theta_{12})\varpi(\theta_{23}) +
 \varpi(\theta_{12})\varpi(\theta_{31}) + 
 \varpi(\theta_{23})\varpi(\theta_{31})\right]$, 
over quasilinear (QL) 
scales have yielded results which appear to agree loosely
with the 
predictions of the so-called hierarchical model (Peebles 1975; Peebles 1980)
for gravitational evolution of Gaussian initial conditions 
(Peebles \& Groth 1975; Fry \& Seldner 1982;
Jing \& Zhang 1989; T\'{o}th, Holl\'{o}si, \& Szalay 1989; 
Borgani, Jing, \& Plionis 1992;
Gazta\~{n}aga 1994; Bernardeau 1995;
Cappi \& Maurogordato 1995; Gazta\~{n}aga, Croft, \& Dalton
1995; Gazta\~{n}aga \& Bernardeau 1998; Jing \& B\"{o}rner 1998;
Magliocchetti \etal 1998).
It is obvious, however, that improved measurements [from surveys such
as the Sloan
Digital Sky Survey (SDSS; Loveday \etal 1998)] 
and more sophisticated theoretical
modeling of angular CFs are required. In particular, 
there has been little theoretical work to date
on the detailed behavior of angular CFs beyond the two-point function, 
despite the practical advantages they can offer.

In this paper, we present a calculation of
the full angular 3PCF\footnote{Frieman \& Gazta\~{n}aga (1999),
using a different method,
have independently derived results for $Z$.
We have compared our
results with theirs and find that they agree for some of the
models we have considered.}, $Z$, as well as
$s_3$, for an arbitrary,
biased, tracer-mass distribution in flat and open cold-dark-matter 
(CDM) models, 
assuming Gaussian initial conditions. The calculation, based
on leading-order PT results, is restricted
to the QL regime, where $\varpi \ll 1$, and further invokes
the small-angle approximation.
We take into account such 
factors as the explicit dependence on the
cosmological parameters, 
the shape and linear evolution
of the CDM power spectrum, and the form of redshift distribution function.
We consider both the redshift distribution of
the APM Galaxy Survey, as well as that of a survey with a mean redshift
of $\overline{z} \simeq 1$.
We also examine the effect of linear and nonlinear bias, which, 
through an extension of the Fry (1996) bias-evolution
model to the case of an arbitrary expansion history,
are allowed to evolve in time. Following BKJ, we 
define the bias parameters, $b_i$, by an expansion of
the unsmoothed tracer-mass field in terms of the unsmoothed dark-matter
field, rather than by relating the smoothed fields, as is often
done (e.g., Fry \& Gazta\~{n}aga 1993). 
While many of the results we obtain 
are similar to those found in a related paper 
(Buchalter \& Kamionkowski 1999, hereafter BK99) dealing with
the spatial 3PCF and skewness, understanding
the detailed behavior of angular
statistics in practice can be equally, if not more, illuminating.
We extend previous derivations of $s_3$ (Bernardeau 1995) to include
a scale-dependent, leading-order correction which arises 
in the case of nonlinear bias, as defined for the unsmoothed fields,
and which becomes
large for positive effective power-spectrum indices. This
behavior, in principle, may allow 
better constraints on the linear- and nonlinear-bias 
parameters on the basis of large-scale skewness measurements alone,
or at least differentiate between the smoothed and unsmoothed biasing scenarios.
For plausible models, we find that $s_3$ is relatively insensitive
to the adopted cosmology, as compared with its stronger dependence
on the biasing scheme. In general, the presence of linear
bias tends to flatten and reduce the scale dependence of $s_3$, while
a nonlinear bias tends to produce a relative increase.
The full angular 3PCF, $Z$, shows similar sensitivities to
the bias parameters, but can also
depend significantly on the cosmological model, and in particular on
the shape of the power spectrum.
We find that the configuration dependence of $Z$ is 
in general more complex than the simple
hierarchical model suggests, and, since angular statistics 
probe clustering over a range of redshifts,
is also sensitive to evolutionary effects.
These properties can 
be used to discriminate between various models which can yield
nearly degenerate predictions for $s_3$, or for the
spatial 3PCF at $z=0$, such as evolving versus
non-evolving bias models, or open versus flat cosmological models. 
Neglecting the full geometric and evolutionary variation of $Z$ throws 
away valuable information which might, in practice, be harder to obtain from
three-dimensional data. In several
instances, we illustrate how accounting for time dependences in, e.g., the
power spectrum or the bias, can impact the resulting calculations,
especially in the case of high-redshift surveys. 
Our theoretical results, in general, should be comparable
to current and future measurements of angular clustering.

In \S2 we outline
the calculations used and present results for the normalized angular skewness.
In \S3 we derive predictions for the full angular 3PCF and in \S4
we summarize our conclusions. Since the calculations are greatly
simplified by working initially in Fourier space, we first 
obtain the angular bispectrum---the two-dimensional
projection of the spatial bispectrum---and then derive the angular 3PCF from
this. A derivation of the equation that relates the spatial and angular
bispectra is presented in the Appendix. This result is essentially a
Fourier-space generalization of Limber's equation for the 3PCF.

\section{ANGULAR SKEWNESS}

Since angular CFs involve
projections of their full spatial
counterparts, our derivations must involve integrals over
the line of sight, which will in general depend on the adopted cosmological 
model.
We shall consider open and flat
Friedmann-Robertson-Walker (FRW) cosmologies with a possible
cosmological constant, so that 
the scale factor,
$a(t)$, satisfies the Friedmann equations,
\begin{eqnarray}
\frac{\dot{a}}{a}  =  H_0 E(z) & \equiv & H_0 \sqrt{\Omega_0 (1+z)^3 + 
(1-\Omega_0 - \Omega_\Lambda) (1+z)^2 + \Omega_\Lambda}, \nonumber \\
\frac{\ddot{a}}{a} & = & H_0^2 [\Omega_\Lambda - \Omega_0 (1+z)^3/2],
\end{eqnarray}
where $\Omega_0$ is the present nonrelativistic matter density
in units of the critical density, $\Omega_\Lambda$ is the contribution
of the cosmological constant to the total present energy density, $H_0$ is the
present value of the Hubble parameter,
and a dot denotes a derivative with respect to time. 
The scale factor is chosen such that
$a_0 H_0 = 2$, where $H_0 = 100$ $h$ km s$^{-1}$ Mpc$^{-1}$. 
If we take our position as the origin, $\bfw=0$, then the angular-diameter
distance to an object at redshift $z$ is given by
\begin{equation}
w(z) = \frac{S(a_0 H_0 f(z) \sqrt{\abso{1-\Omega_0-\Omega_\Lambda}})}
{a_0 H_0 \sqrt{\abso{1-\Omega_0-\Omega_\Lambda}}}; \;\;\;\;\;\;
f(z) = \frac{1}{2} \int_{0}^{z} \frac{dz^{\prime}}{E(z^{\prime})}, 
\label{w}
\end{equation}
where $S(x) = \{\sinh{x}$, $x$, $\sin{x}\}$, respectively,
for open, flat, and closed geometries.
For an Einstein-de Sitter universe, $w(z) = f(z)$. The distance
to the horizon is given by $\eta_0 = w(\infty)$.

We wish ultimately to derive an expression for the angular 3PCF
of a distribution of tracer masses (e.g., galaxies, quasars,
clusters, radio sources,
etc.), $Z({\theta}_{12},{\theta}_{23},{\theta}_{31})=
\VEV{{p}(\bftheta_1){p}(\bftheta_2){p}
(\bftheta_3)}$, where $\bftheta$ represents two-dimensional coordinates
on the sky, $p(\bftheta) = [\Sigma(\bftheta)-\overline{\Sigma}]/
\overline{\Sigma}$ is the
fractional perturbation
to the unsmoothed tracer-mass density field,  $\Sigma(\bftheta)$,
and angular brackets denote an average over direction in the sky, with
fixed values for the distances $\theta_{ij} = \abso{\bftheta_i - \bftheta_j}$. 
Though statistical homogeneity and isotropoy ensure that the 3PCF will
only depend on the relative angular separations between the three points,
measuring $Z$ for all possible geometric configurations
can nonetheless be a daunting task.
For this reason, many studies have focussed instead on the
normalized angular skewness, $s_3 \equiv \VEV{p_{\Theta}^3(\bftheta)}/
\VEV{p_{\Theta}^2(\bftheta)}^2$, obtained from the moments
of counts in cells smoothed with an effective angular scale half-width
$\Theta$ (it is implicitly assumed that $\Sigma
\propto n$, where $n$ is the number
of discrete counts).
This one-point statistic is more readily calculable than
the full angular 3PCF, while still preserving
information about the overall scale-dependence of $Z$.
 
In deriving the following results, 
we implicitly make use of the small-angle approximation
in assuming that a small patch of sky can be treated using a Fourier expansion
(Peebles 1980). In this approximation, the component of the 
three-dimensional wave vector along the 
line of sight is taken to be negligible 
compared to the orthogonal components, so that
the angular CFs depend only on the latter. We later comment
on this assumption.
We define $p_{\Theta}$ using a two-dimensional spherical top
hat window function, so that in terms of the Fourier components,
$\tilde{p}(\bfkappa)$,
of the unsmoothed two-dimensional density field, we have
\begin{equation}
p_{\Theta}(\bftheta) = \int \frac{d^{2}\bfkappa}{(2\pi)^2}
\tilde{p}(\bfkappa) e^{i \bfkappa\cdot\bftheta} {\cal{W}}(\kappa\Theta),
\end{equation}
where $\bfkappa$ is a two-dimensional wave vector with magnitude
$\kappa$, $\W2(x) = 2J_1(x)/x$ is the Fourier transform of the 
two-dimensional spherical top-hat
window function, and $\Theta$ is an angular smoothing radius.
We can then obtain the angular 2PCF, as measured using counts in cells,
\begin{equation}
\varpi_{\Theta}(\theta_{12})=\VEV{p_{\Theta}(\bftheta_1)
p_{\Theta}(\bftheta_2)} = 
\int\!\int
\frac{d^2\bfkappa_1}{(2\pi)^2}\frac{d^2\bfkappa_2}{(2\pi)^2}
e^{i(\bfkappa_1\cdot\bftheta_1 + \bfkappa_2\cdot\bftheta_2)} 
\W2(\k_{1}\Theta) \W2(\k_{2}\Theta)
\VEV{\tilde{p}(\bfkappa_1)\tilde{p}
(\bfkappa_2)}.
\label{varpi}
\end{equation}
We define 
\begin{equation}
\VEV{\tilde{p}(\bfkappa_1)\tilde{p}
(\bfkappa_2)} \equiv (2\pi)^2 \delta_D(\bfkappa_1+\bfkappa_2) P_p(\kappa),
\label{Pp}
\end{equation}
where the angular, or projected, power spectrum of the tracer mass, $P_p$
(usually written 
in terms of $C_\ell$'s in papers on the cosmic microwave background), 
can be related to the full tracer-mass power spectrum, $P(k,w)$, 
by a Fourier-space analog of Limber's
equation (Kaiser 1992),
\begin{equation}
P_p = \int_0^{\eta_0} \frac{dw}{w^2} \left( \frac{dN}{dw} \right)^2 P(\k/w,w).
\label{Pp2}
\end{equation}
Here, $dN/dw$ is a $w$-space selection function for the tracer mass, 
normalized such that $\int_0^{\eta_0} dw (dN/dw)
= 1$. 

We assume the fractional perturbation to the {\em unsmoothed}
three-dimensional tracer mass density field, $\delta(\bfr)$, 
may be expanded in terms of the local perturbation to 
the unsmoothed, underlying matter field, $\delta_m(\bfr)$, via
\begin{equation}
     \delta = b_1 \delta_m + \frac{b_2}{2} \delta_m^2 + \cdots,
\label{dexp}
\end{equation}
where $b_1$ is the linear bias term, $b_2$ the first nonlinear
term, etc., and $\delta_m$ is itself written as $\delta_m = 
\delta_{m}^{(1)} + \delta_{m}^{(2)} + \cdots,$ 
where $\delta_{m}^{(n)} \ll \delta_{m}^{(n-1)}$, $\delta_{m}^{(1)}$ 
is the linear
solution, and $\delta_{m}^{(2)}$ characterizes the leading-order 
departure from the Gaussian initial conditions. 
The linear solution for the underlying spatial density contrast has the 
separable form
\begin{equation}
\delta^{(1)}_m(\bfr,w) = D(w) \delta^{(1)}_m(\bfr,0), 
\label{dm1}
\end{equation}
so that fluctuations evolve simply as the linear growth factor, $D(w)$.
This in turn implies that, to leading order, the spatial
and time dependence of the power spectrum can likewise be factorized.
For FRW cosmologies, the growth factor is given
(as a function of redshift) by
\begin{equation}
D(z) = \frac {5 \Omega_0 E(z)}{2} \int_z^{\infty} dz^{\prime} \frac
{1+z^{\prime}}{\left[E(z^{\prime})\right]^3}.
\label{D}
\end{equation}
For an Einstein de-Sitter universe $D(w)$ is simply the scale
factor, $a(w)$.
Assuming the unsmoothed three-dimensional 
tracer-mass density contrast to be related 
to the underlying distribution via equation~(\ref{dexp}), we 
can write the leading-order result
for the full linear power spectrum of the tracer mass,
\begin{equation}
P(k,w) = A b_1^2 D^2(w) k^n T^2(k),
\label{Pkt}
\end{equation}
where $k^n$ is the primordial power spectrum,
$A$ is the overall amplitude, and $T(k)$ is a model-dependent transfer 
function.
Substituting equations~(\ref{Pkt}), 
(\ref{Pp2}), and (\ref{Pp}) into (\ref{varpi}), 
taking $\bftheta_1 = \bftheta_2 \equiv
\bftheta$, and
$x=\k\Theta$, we arrive at the angular variance (the 2PCF at zero lag),
\begin{equation}
\varpi_{\Theta}(0)=\VEV{p_{\Theta}^{2}(\bftheta)} = \frac{A }
{2\pi \Theta^{n+2}}
\int_0^{\eta_0} dw \left( \frac{dN}{dw} \right)^2 \frac{{b_1}^2 D^2(w)}{w^{n+2}}
\int_0^{\infty} dx \: x^{n+1} \W2^2(x) T^2(x/\Theta w),
\label{varpizero}
\end{equation}
where the $b_1$ term is left inside the $w$ integral to allow for the
possibility of bias evolution.

The counts-in-cells angular 3PCF is given by
\begin{eqnarray}
Z_{\Theta}(\theta_{12},\theta_{23},\theta_{31}) & = & \VEV{{p_{\Theta}}
(\bftheta_1){p_{\Theta}}(\bftheta_2){p_{\Theta}}
(\bftheta_3)} = \int\!\int\!\int
\frac{d^2\bfkappa_1}{(2\pi)^2}\:
\frac{d^2\bfkappa_2}{(2\pi)^2}\:\frac{d^2\bfkappa_3}{(2\pi)^2}
e^{i(\bfkappa_1\cdot\bftheta_1 + \bfkappa_2\cdot\bftheta_2 + 
\bfkappa_3\cdot\bftheta_3)} \nonumber \\
&   & \times \VEV{\tilde{p}(\bfkappa_1)\tilde{p}
(\bfkappa_2)\tilde{p}(\bfkappa_3)}  
\W2(\k_{1}\Theta) \W2(\k_{2}\Theta)
\W2(\k_{3}\Theta).
\label{askew}
\end{eqnarray}
The angular, or projected, bispectrum, $B_p$, is defined via
\begin{equation}
\VEV{\tilde{p}(\bfkappa_1)\tilde{p}(\bfkappa_2)\tilde{p}
(\bfkappa_3)} \equiv (2\pi)^2 \delta_D(\bfkappa_1+\bfkappa_2+\bfkappa_3)
B_p(\k_1,\k_2,\k_3),
\label{Bp1}
\end{equation}
and is related to the full three-dimensional bispectrum, $B$, by
\begin{equation}
B_p(\k_1,\k_2,\k_3) = \int_{0}^{\eta_0} \frac{dw}{w^4}
\left( \frac{dN}{dw} \right)^3 B(\k_1/w,\k_2/w,\k_3/w,w)
\label{Bp2}
\end{equation}
(see Appendix).
The full bispectrum is given, to leading order in PT, by 
\begin{eqnarray}
    B({k}_{1},{k}_{2},{k}_{3},w) & = & 
    P(k_{1},w)P(k_{2},w)  \Biggl\{ \frac{1}{b_1}
    \left[ 1+\mu + \cos\psi\left(\frac{k_{1}}{k_{2}} + 
    \frac{k_{2}}{k_{1}}\right)
    + (1-\mu)\cos^{2}\psi \right]  \nonumber \\ 
    &   & + \frac{b_2}{{b_1}^2} \Biggr\} \;\; + \;\; \mbox{(cyc.)},
\label{Bf2}
\end{eqnarray}
(Fry 1984; Goroff \etal 1986;
Matarrese, Verde, \& Heavens 1997)
where $\psi$ is the angle between $\bfk_{1}$ and $\bfk_{2}$, and $\mu$ 
is a function of the expansion history, equal to $3/7$ for
an Einstein-de Sitter universe (Peebles 1980)
and differing from this value only
slightly for other reasonable choices of the density parameters
(Bouchet \etal 1992; Bernardeau 1994; 
Bouchet \etal 1995; Catelan \etal 1995; Martel 1995; Scoccimarro
\etal 1998; Kamionkowski \& Buchalter 1998). 
Note that the expression for
$B$ includes the dependence on $b_2$, since the leading-order result
for the 3PCF includes second-order terms in the expansion of 
equation~(\ref{dexp}).

Evaluating equation~(\ref{askew}) with
$\bftheta_1 = \bftheta_2 = \bftheta_3 \equiv \bftheta$,
$x_i = \k\Theta$, taking $\bfkappa_1$ to lie in the $\psi=0$ direction,
and substituting equations~(\ref{Bp1}), (\ref{Bp2}), and (\ref{Bf2}) 
we obtain, 
after a little algebra, an expression
for the area-averaged skewness, 
\begin{eqnarray}
\VEV{p_{\Theta}^{3}(\bftheta)} & = & \frac{3 A^2}
{(2\pi)^3 \Theta^{2n+4}}
\int_{0}^{\eta_0} dw \left( \frac{dN}{dw} \right)^3 \frac{ {b_1}^4 
D^{4}(w)}{w^{2n+4}} \nonumber \\
&\times& \int_{0}^{\infty} dx_1 x_1^{n+1} \W2(x_1) T^{2}(x_1/\Theta w)
\int_{0}^{\infty} dx_2 x_2^{n+1} \W2(x_2) T^{2}(x_2/\Theta w) \nonumber \\
                    &\times& \int_{0}^{2\pi} d\phi \: \W2(x_3) \left( 
\frac{1}{b_1} \left[(1+\mu) + \cos\psi(\frac{x_1}{x_2} + \frac{x_2}{x_1}) + 
(1-\mu)\cos^{2}\psi \right] + \frac{b_2}{{b_1}^2} \right),
\label{askewness}
\end{eqnarray}
where one integral vanishes under the requirement $\bfkappa_3 =
- (\bfkappa_1 + \bfkappa_2)$, and
a factor of 3 arises from symmetry considerations applied to the
2 cyclic permutations in equation~(\ref{Bf2}). 
Noting that $x_3 = \sqrt{x_1^2 + x_2^2 +
2x_1 x_2 \cos\psi}$, we can evaluate the $\psi$ integrals
by using the
summation theorems for Bessel functions (Gradshteyn \& Ryzhik 1980;
Bernardeau 1995) 
which yield:
\begin{eqnarray}
\int_0^{2\pi} d\psi \: \sin^2 \psi \: \W2(x_3) & = & \pi \W2(x_1) \W2(x_2), \\
\int_0^{2\pi} d\psi \: (1 + \frac{x_2}{x_1}\cos\psi) \: \W2(x_3) & = & 2\pi
 J_0(x_2) \W2(x_1), \\
\int_0^{2\pi} d\psi \: \W2(x_3) & = & 2\pi \sum_{j=0}^{\infty} (2j+1) 
\W2_{2j+1}(x_1)\W2_{2j+1}(x_2),
\label{2ds}
\end{eqnarray}
where $\W2_n(x) = 2J_{n}(x)/x$, so that $\W2(x) = \W2_1(x)$.
Using the above results gives
\begin{eqnarray}
\VEV{p_{\Theta}^{3}(\bftheta)} & = & \frac{3 A^2}{(2\pi)^2 
\Theta^{2n+4}}
\int_{0}^{\eta_0} dw \left( \frac{dN}{dw} \right)^3 \frac{ {b_1}^4 D^{4}(w)}{w^{2n+4}}
\nonumber \\
&   & \int_{0}^{\infty} dx_1 x_1^{n+1} \W2(x_1) T^{2}(x_1/\Theta w)
\int_{0}^{\infty} dx_2 x_2^{n+1} \W2(x_2) T^{2}(x_2/\Theta w) \nonumber \\
&   & \Biggl\{ \frac{1}{b_1} \left[ J_0(x_2)\W2(x_1) + J_0(x_1)\W2(x_2) \right]
- \frac{1}{2}\frac{(1-\mu)}{b_1} \W2(x_1)\W2(x_2)  \nonumber \\
&   & +  \frac{b_2}{b_1^2} \sum_{j=0}^{\infty} 
(2j+1)\W2_{2j+1}(x_1)\W2_{2j+1}(x_2) \Biggr\} .
\label{askewness2}
\end{eqnarray}

We re-emphasize that our results in general are based on 
leading-order PT and are thus restricted to the 
QL regime, i.e., scales large enough so that the rms density contrast
fluctuations are small compared with unity.
Numerical simulations and observations both confirm that 
higher-order nonlinear corrections have little impact
on spatial statistics on
QL scales (Szalay 1988; T\'{o}th, Holl\'{o}si, \& Szalay 1989;
Gott, Gao, \& Park 1991; Fry, Melott, \& Shandarin 1993; Jain \& 
Bertschinger 1994;
Scoccimarro \etal 1998). By comparing with
$N$-body results, Gazta\~{n}aga \& Bernardeau (1998)
test the validity of PT as applied to angular statistics and find
good agreement on scales $\Theta \ga 1^{\circ}$, with the details depending
on the shape of the power spectrum.
In addition, simulations show  
that QL scales can still obey 
leading-order PT even when smaller scales have become fully nonlinear 
(Bouchet \& Hernquist 1992),
and further that the predictions of QL PT may hold even on 
scales where the rms fluctuation 
exceeds unity (Bernardeau 1994; Baugh, Gazta\~{n}aga, \& Efstathiou 1995;
Fry, Melott, \& Shandarin 1995; Szapudi \etal 1998).

Our results have also relied on the use of the small-angle approximation.
The Fourier integrals used to calculate $\varpi$ and $Z$, however, extend
over all values of $\kappa$, including small values for which this
approximation breaks down, but which nonetheless may contribute significant
weights to the integrands on degree scales.\footnote{We note 
that this difficulty
could be avoided if one considers the bispectrum directly.} 
Moreover, the small-angle approximation
might appear to go against the assumption of quasilinearity; in practice, the
degree to which these two assumptions are mutually plausible will
in general depend on the selection function in question.
Despite these considerations, 
our results for $s_3$ and $q$ on scales near 
$1^{\circ}$ do agree well with data from observations 
(e.g., Gazta\~{n}aga 1994; see Figure \ref{fig:s3_apm}) 
and $N$-body simulations
(Frieman \& Gazta\~{n}aga 1999) on these scales.
Bernardeau (1995) investigates the validity
of the small-angle approximation for models with different
redshift selection functions, $dN/dz$,
and finds good agreement with numerical results for $s_3$
on scales near and below $1^{\circ}$.
Gazta\~{n}aga \& Bernardeau (1998) find that the small-angle approximation
can yield reasonable agreement with numerical
results for $s_3$ out to the $5^{\circ}-10^{\circ}$ scale. Verde \etal (1999)
perform an exact calculation for the angular 3PCF using a  
spherical-harmonic decomposition, and a more complete
comparison and assessment of the small-angle results against the exact results
are presented therein. 

\subsection{Effects of the Power Spectrum}

Equations~(\ref{varpizero}) and (\ref{askewness2}) are 
one-point results which, 
for a given choice of $\Omega_0$, $\Omega_\Lambda$, $b_1$, and $b_2$, 
depend only on the smoothing radius, $\Theta$, 
and the form of the power spectrum.
If we assume, for computational simplicity, a scale-free power spectrum,
$P(k) \propto k^n$ [$T(k)=1$], then
we can calculate an expression for $s_3$ where the $x$ integrals
can be separated from the $w$ integration and evaluated analytically;
combining equations~(\ref{varpizero}) and
(\ref{askewness2}), we obtain
\begin{equation}
s_3 = \frac{\VEV{p_{\Theta}^3(\bftheta)}}
{\VEV{p^2_{\Theta}(\bftheta)}^2}
= R_3(n) \left\{\frac{1}{b_1}\left[\frac{36}{7} + \frac{9}{14}
\left(\frac{7}{3}\mu-1\right)
- \frac{3}{2}(n+2) \right] + 3\frac{b_2}{b_1^2} \left[ 1 + \Delta(n)
\right] \right\},
\label{s3}
\end{equation}
where
\begin{equation}
R_3(n) = \frac{\int_0^{\eta_0} dw \left( dN/dw \right)^3 
w^{-(2n+4)}D^4(w)}{\left[ \int_0^{\eta_0} dw \left( 
dN/dw \right)^2 w^{-(n+2)} D^2(w) \right]^2},
\end{equation}
and
\begin{equation}
\Delta(n) = \frac {\sum_{j=1}^{\infty} (2j+1) \left[ \int_0^{\infty}
dx \: x^{n-1} J_1(x) J_{2j+1}(x) \right]^2}{\left[ \int_0^{\infty} 
dx \: x^{n-1} J_1^2(x) \right]^2}.
\label{Delta}
\end{equation}
\begin{figure}[htbp]
\plotone{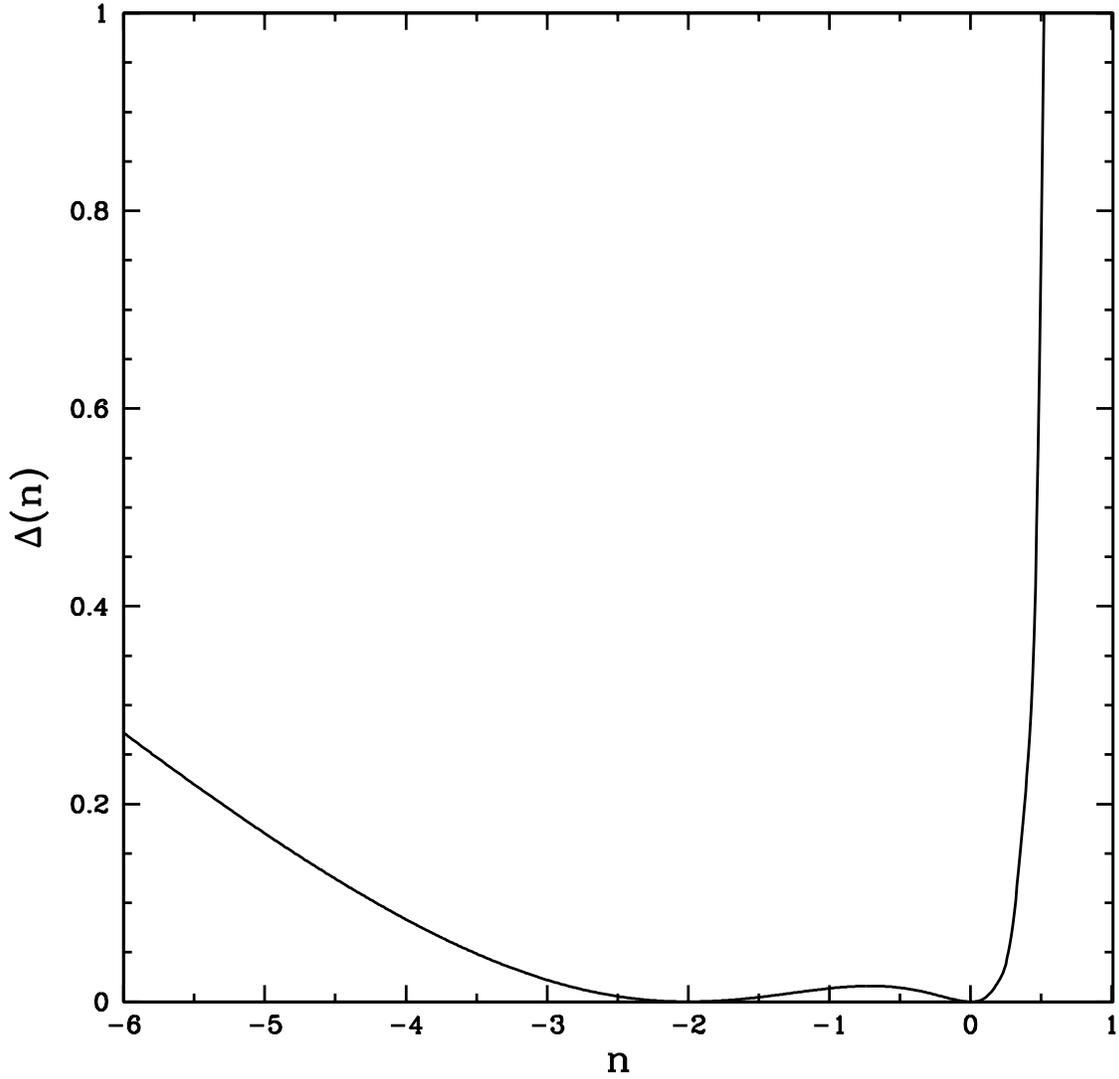}
\caption{The variation of the $\Delta$ term in equation~(\ref{s3}) with
spectral index $n$. Though negligible for $n$ in the range $-2<n<0$, this
term rapidly becomes significant for $n>0$, due to the impact of
small-scale fluctuations.}
\label{fig:Delta}
\end{figure}
Note that while the term in curly brackets in equation~(\ref{s3})
depends only very weakly on $\Omega_0$ through $\mu$, the $R_3(n)$
term depends on the cosmological model both through the presence of the
linear growth factor, $D$ (which is often neglected), as well as through
the explicit dependence of $w$ on the expansion history 
[see equation~(\ref{w})].
Equations~(\ref{s3})--(\ref{Delta}) are valid for $n$ in the range $-2>n>1$; 
the individual terms in the
series for $\Delta(n)$ can be evaluated explicitly (Watson 1966) and 
we find that $\Delta(n) \ll 1$
for $-2<n<0$, but diverges for $n>0$, as illustrated in Figure \ref{fig:Delta}.
 
The bracketed $n$-dependent terms in equation~(\ref{s3}) arise from the effects
of smoothing the contiguous field, $\delta(\bfr)$, defined in 
equation~(\ref{dexp}). In particular, since smoothing at a fixed
physical mixes different mass scales, the higher-order moments for smoothed
fields will generally contain additional terms reflecting the scale dependence
of the rms fluctuation (Bernardeau 1994, 1995).
The $n=-2$ case thus recovers the no-smoothing result for $s_3$
(Bernardeau 1995), since the rms fluctuation in this case is independent
of scale [see equation~(\ref{varpizero})].
Though the filter is designed to separate out
the nonlinear, small-scale fluctuations, an unbounded scale-free power
spectrum with $n>0$ produces so much small-scale power that
the fluctuations induced by smoothing over nonlinear scales become 
comparable to
or greater than the large-scale linear perturbations, leading to the divergence
in $\Delta(n)$ seen in the Figure.

Equation~(\ref{s3}) differs from the known 
result for $s_3$ assuming constant, nonlinear bias
(Bernardeau 1995) only in that it contains the $\Delta(n)$ term.
The difference is attributable to the fact 
that previous authors usually define the bias parameters
using the form of equation~(\ref{dexp}), but relating
the {\em smoothed}, rather than the unsmoothed, observed
and underlying density fields. Such a definition
is technically nonlocal, only fixing the linear- and nonlinear-bias 
parameters at the chosen smoothing scale. While it can be shown
that the ``smoothed'' and ``unsmoothed'' linear-bias parameters
are identical, the value of the smoothed nonlinear-bias parameter
will, unlike the unsmoothed $b_2$ defined in equation~(\ref{dexp})
in the limit of continuous fields, depend on the smoothing scale.
Comparing the result for $s_3$ with those of Fry \& Gazta\~{n}aga (1993)
and Bernardeau (1995), 
we can infer that the smoothed nonlinear-bias parameter will vary
as $b_2 [1+\Delta(n)]$. In the limit of no smoothing ($n=-2$), the two
parameters are identical, as expected, but can differ dramatically
on large scales (see \S2.2). Similar correction
terms, arising from nonlinear bias, are expected for
higher-order angular moments as well.
Such terms might be used to 
discriminate between
the smoothed and unsmoothed bias pictures, and possibly
distinguish between linear and nonlinear bias
based on the skewness alone, as discussed in \S2.2 and in BK99.
Unless stated otherwise, it will be hereafter understood that ``bias''
shall refer to the unsmoothed bias prescription defined by 
equation~(\ref{dexp}).

While the power spectrum for any viable cosmological model will not be
given by a simple power law, it can be shown, using the results
of Bernardeau (1995) and Gazta\~{n}aga \& Bernardeau (1998), that
an exact result for $s_3$ can be obtained for arbitrary power spectra,
using a properly redefined {\em effective} index (Scoccimarro 1998). 
This result, analogous to that in the spatial case (Bernardeau 1994),
is strictly valid, however, only for linear biasing, since it fails to
account for the above-mentioned large-scale variation associated with $b_2$. 
Since we wish to consider nonlinear biasing, and
because, unlike $s_3$, no corresponding exact result can be obtained for the
full angular 3PCF in the case of arbitrary power spectra, 
even for linear biasing, the results of this paper are based entirely on
numerical integration of the appropriate equations.
Many of these will involve highly oscillatory
integrands, making them difficult to evaluate; we have performed
checks on the numerical
accuracy of our results, and found them to be good to 
within a few percent.

For realistic CDM models, we employ a transfer
function given by
\begin{equation}
T(q) = \frac{\ln (1+2.34q)/(2.34q)}{\left[ 1+3.89q + (16.1q)^2
+(5.46q)^3 + (6.71q)^4 \right] ^{1/4}}
\label{Tk}
\end{equation}
(Bardeen \etal 1986), where $q=k_p/ \Gamma$, $\Gamma \approx \Omega_0 h$,
and $k_p$ is the physical wavenumber
in units of $h$ Mpc$^{-1}$, related to the comoving wavenumber, $k$,
through our adopted normalizations by 
$k_p = k/6000$ $h$ Mpc$^{-1}$. The result for $s_3$ is then 
obtained by using equation~(\ref{Tk}) together with (\ref{varpizero}) and
(\ref{askewness2}). Note that in the case of a scale-dependent transfer
function the $x$ and $w$ integrals cannot be separated.
\begin{figure}[htbp]
\plotone{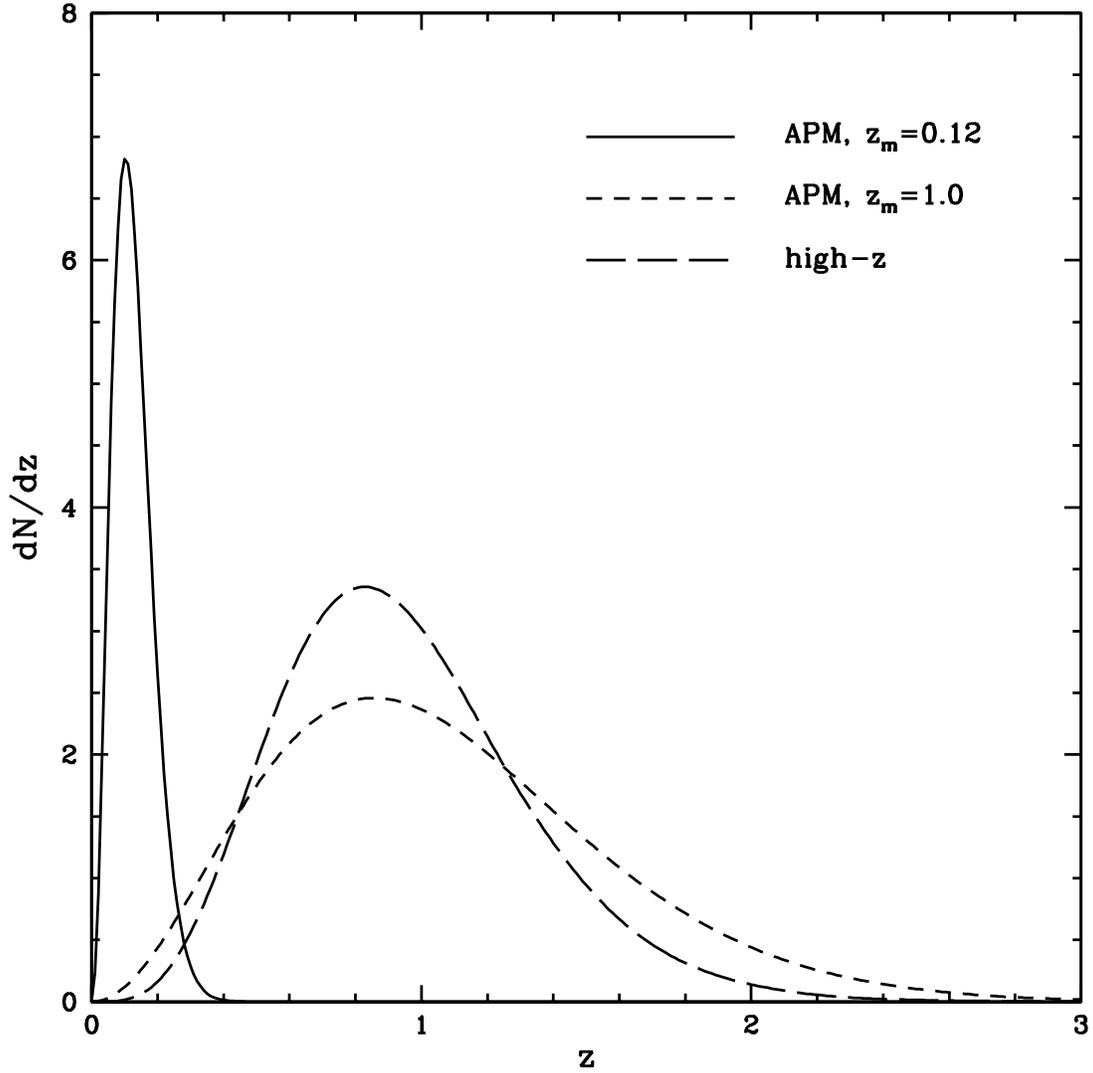}
\caption{Redshift-space plot of the APM 
($z_m=0.12$) and high-$z$ ($\overline{z} \simeq 1$)
selection functions employed in the calculations. For comparison,
we also display the APM SF assuming $z_m=1.0$; both this and the high-$z$
SF have been assigned three times the normalization of the low-redshift
APM SF for the purposes of representation.}
\label{fig:dndz}
\end{figure}

Our calculations also require the assumption of a selection function (SF)
along the line of sight, and we consider two functional forms
chosen to characterize low-redshift ($\overline{z} \sim 0.1$)
and high-redshift surveys ($\overline{z} \sim 1$). 
For the former, we consider the SF of the Automatic
Plate Measuring (APM) Galaxy Survey,
\begin{equation}
\frac{dN}{dz} \propto z^2 \exp\left[ - \left( \frac{z}{z_c}\right)^{3/2}\right];
\:\:\:\:\:\:\:\: z_m = 1.412 z_c,
\end{equation}
where $z_m$ is the median reshift of the survey, taken to be 0.12. 
This functional form provides very good
fit to the APM redshift distribution over the entire APM magnitude range
(Baugh \& Efstathiou 1993; Maddox, Efstathiou, \& Sutherland 1996;
Gazta\~{n}aga \& Baugh 1998), and should also approximate comparable
optical surveys, such as the SDSS. 
To compare the expected results for a survey
such as the APM, which reaches only modest redshifts ($z < 0.4$),
with those from a much deeper angular survey, we also employ the
$w$-space SF of Kaiser (1992) (intended to mimic 
a magnitude-limited survey),
\begin{equation}
\frac{dN}{dw} = \frac{\beta w^{\alpha}\exp\left[-(w/w_{*})^{\beta}\right]}
{w_*^{1+\alpha} \Gamma\left[(1+\alpha)/\beta\right]}; \:\:\:\:\:\:\:\:
w_* = \overline{w}\left(\Gamma\left[(2+\alpha)/\beta\right]\Gamma
\left[(1+\alpha)/\beta\right]\right),
\label{kaiser}
\end{equation}
where $\overline{w}$ is the mean conformal lookback time. 
We will hereafter choose $\alpha=4$, $\beta=4$, and
$\overline{w}=0.35$, which yields a peak redshift of $z \sim 0.8$,
so that this model, which we will refer to as the ``high-$z$'' SF,
might characterize, for example, a radio survey such as the
VLA FIRST Survey (Becker, White,
\& Helfand 1995),
with a median redshift of $z_m \sim 1$ (Cress \& Kamionkowski 1998).
Figure \ref{fig:dndz} shows the above APM SF
together with the high-$z$ SF, as well as
the APM SF obtained using $z_m = 1.0$, for comparison. 
Baugh \& Efstathiou (1993) and Maddox, Efstathiou \& Sutherland (1996)
find that angular statistics in the APM Survey depend much more sensitively
on the peak redshift of the SF, rather than
on its precise shape; we find this conclusion to hold in general
for SFs we consider. Bernardeau (1995) finds that the small-angle
approximation works reasonably well for an APM-like SF; for deeper
surveys in which a fixed angle can probe larger scales,
this approximation is expected to provide still better results.
 
\begin{figure}[htbp]
\plotone{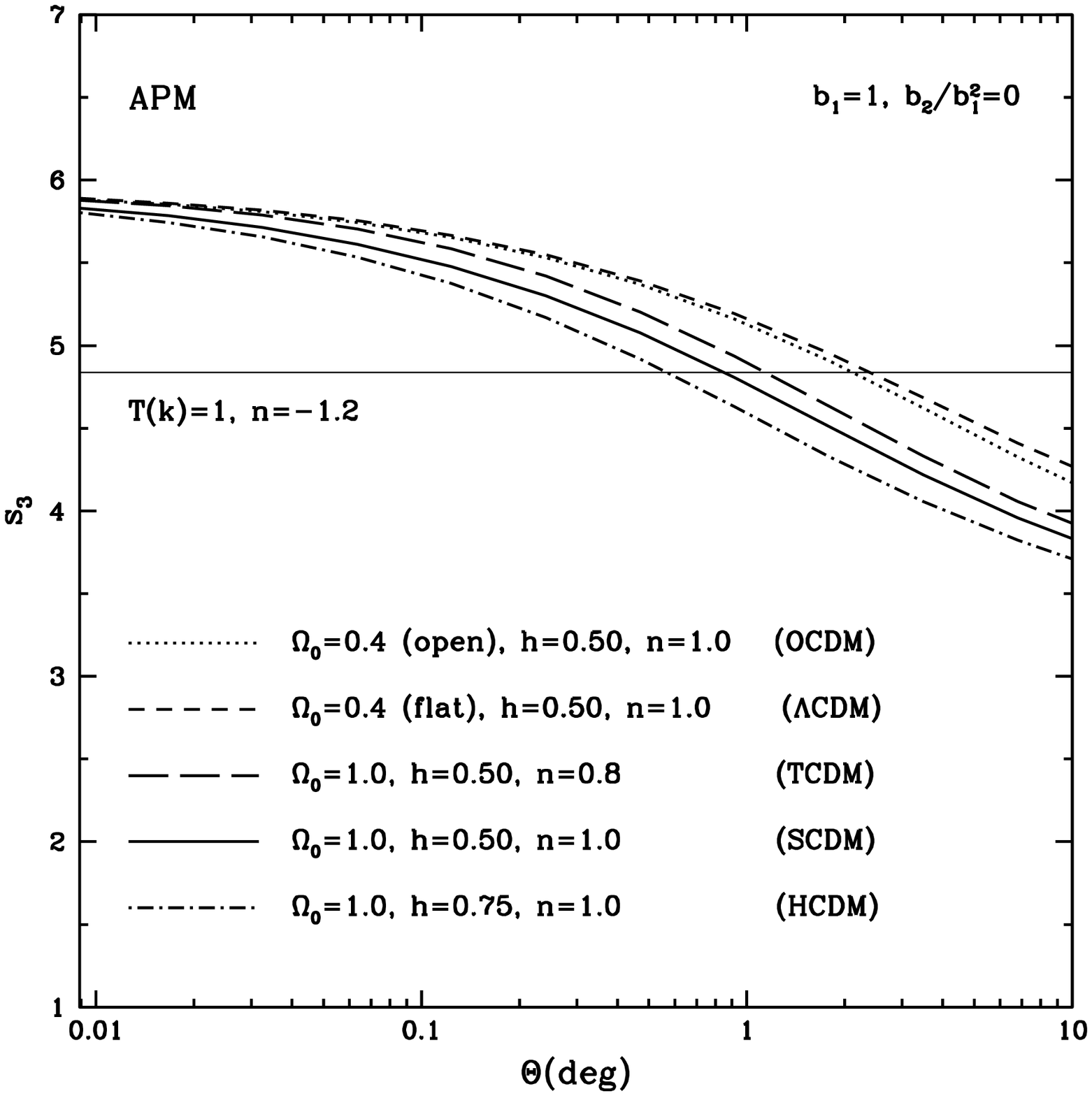}
\caption{The variation of $s_3$ with $\Theta$, for the APM SF, in the five
cosmological models described in the text, assuming no bias. Note that
the differences between these models are relatively small. The large
values of $s_3$ at high $\Theta$ are due to the shallowness of the SF. The
solid, horizontal line shows the result obtained from equation~(\ref{s3})
for a scale-free power spectrum with $n=-1.2$.}
\label{fig:s3_a12}
\end{figure}
\begin{figure}[htbp]
\plotone{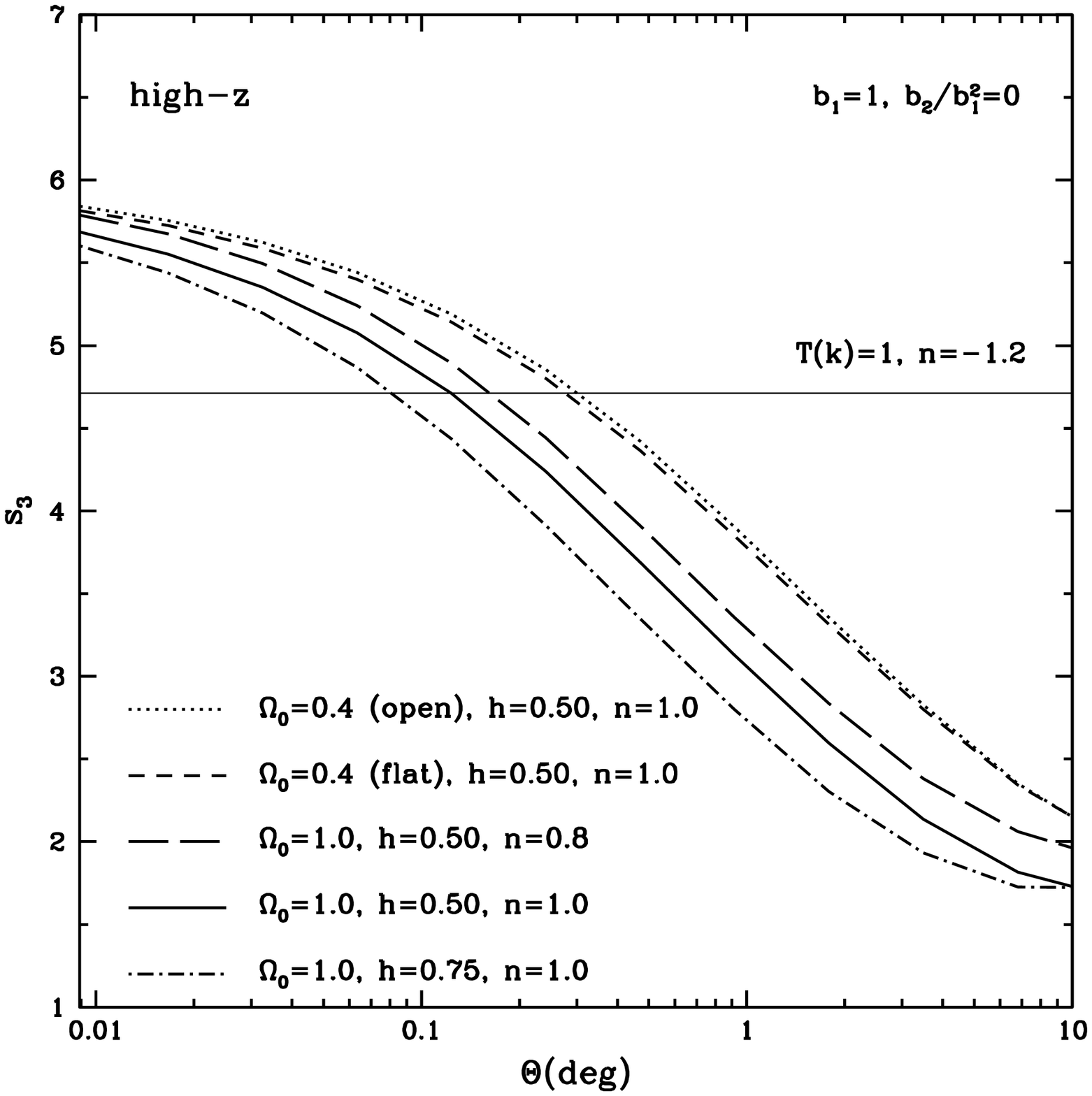}
\caption{Same as Figure \ref{fig:s3_a12}, but using the high-$z$ SF. The models
differ slightly more than in the APM case, and attain smaller values of
$s_3$ at high $\Theta$ since these angular scales, in a deeper survey,
incorporate larger physical scales where the effective power-spectrum
index is larger.}
\label{fig:s3_k}
\end{figure}

In Figure \ref{fig:s3_a12}, we plot the variation of $s_3$ with $\Theta$,
assuming the APM SF, for an unbiased
($b_1 = 1$, $b_2 =0$) tracer-mass population in four flat
($\Omega_0+\Omega_\Lambda=1$) 
cosmological models: standard CDM (SCDM; $\Omega_0=1$, 
$h=0.5$, $n=1$),  
a tilted CDM model (TCDM; $\Omega_0=1$,
$h=0.5$, $n=0.8$), CDM with a cosmological constant 
($\Lambda$CDM; $\Omega_0=0.4$, $h=0.5$, $n=1$), and
CDM with a high
Hubble parameter (HCDM; $\Omega_0=1$, $h=0.75$, $n=1$),
as well as an open CDM model (OCDM; $\Omega_0=0.4$, $h=0.5$, $n=1$).
The thin horizontal line in the Figure shows the (virtually
$\Omega_0$-independent) semi-analytic results
obtained for a scale-free
power spectrum with a canonical index value of $n=-1.2$ (Peebles 1980).
The results for $s_3$ in general are very similar to those found by BK99
for the normalized spatial skewness.
The scale-dependent CDM transfer function
naturally induces a dependence on the
smoothing scale, $\Theta$, and compared to equation~(\ref{s3}),
a more substantial dependence on the combination
of cosmological parameters $\Gamma = \Omega_0 h$.
Yet, although the five cosmological
models shown in the Figure span a broad range of  
parameter values, $0.2 < \Gamma < 0.75$, they yield fairly similar results,
and are
certainly all consistent within the limits of current observations (see 
Figure \ref{fig:s3_apm}).
The slight variation seen is mostly due 
to the dependence on $\Gamma$;
there is little difference between
the SCDM and TCDM models, which differ only in the value of $n$,
and virtually no difference between the OCDM and $\Lambda$CDM models.
If the analysis is restricted to the
range $0.2 < \Gamma < 0.3$, as suggested by current data
(Bartlett \etal 1998), one can infer
from the Figure that the predicted values for $s_3$ for a survey
such as the APM are relatively
insensitive to the adopted cosmological model.    
Figure \ref{fig:s3_k} shows the results for $s_3(\Theta)$ for the same five
cosmological models, assuming the high-$z$ SF.
The primary dependence is again seen to be with $\Gamma$,
and while there is a somewhat greater 
spread between the different models, the differences between 
realistically viable
models are still relatively small compared to those induced
by the bias dependences which we explore in the next section. 

Overall, we find that the high-$z$ SF curves for these models
show greater variation of $s_3$ with $\Theta$ than their
APM SF counterparts.
This is simply due to the fact that in a deeper angular survey,
a fixed angle probes clustering over a broader range of physical 
scales, including very large scales, while
conversely, in a shallower survey, a fixed angle is sensitive
to clustering only over a relatively narrow range of smaller physical scales.
Thus, at larger values of $\Theta$, the high-$z$ result
incorporates physically large scales and tends
toward smaller values for $s_3$. These can be roughly 
approximated using equation~(\ref{s3}) by
$s_3 \approx R_3(n_{k}) \times \left[ 36/7-(3/2)(n_{k}+2) \right] \approx 
R_3(n_{k}) \times (9/14)$, where $n_k$, the
value of the effective spectral index of the (untilted) linear-theory CDM
power spectrum, tends towards unity at large scales 
(recall that $R_3$ is of order 
unity and independent of $\Theta$). For the APM case, however, $s_3$
remains higher at larger $\Theta$, since even large angular scales
in a shallower survey still correspond only to smaller physical scales, where
the effective power-spectrum index is negative. 
At smaller $\Theta$, both sets of curves tend toward
the expected larger values of $s_3 \approx R_3(n_{k}) \times (93/14)$ obtained 
with $n_{k} \approx -3$---the effective power-spectrum index at small scales;
for tilted models, $s_3$ tends toward a value $n_k-1$ smaller.
Note that in both Figures, the semi-analytic results
for a scale-free
power spectrum with a canonical index value of $n=-1.2$ 
(thin, horizontal lines)
provide poor fits to the CDM predictions. 
Other authors have investigated QL results for the normalized 
angular skewness
using CDM models (Gazta\~{n}aga \& Frieman 1994; Frieman \& Gazta\~{n}aga 1994;
Gazta\~{n}aga \& Baugh 1998; Gazta\~{n}aga \& Bernardeau 1998), 
and find similar results.

\subsection{The Effects of Bias and its Evolution}

The results of the previous section already allow for the existence
of a constant, nonlinear bias between the tracer-mass and
the underlying matter distributions,
but can be generalized by allowing
the bias parameters to evolve with time, as is suggested both by theory
and observations (Fry 1996; Peacock 1997; Matarrese \etal 1997;
Steidel \etal 1998; Cress \& Kamionkowski 1998; Catelan \etal 1998;
Taruya, Koyama, \& Soda 1998; Tegmark \& Peebles 1998; Col\'{i}n \etal 1998;
Baugh \etal 1998; Magliocchetti \& Maddox 1998). 
Fry (1996) proposes a
bias-evolution model, which assumes that
objects in an Einstein-de Sitter universe
form at a fixed formation redshift, $z_f$, by some arbitrary local
process which induces a bias at that epoch, and are
subsequently governed purely by gravity. 
We note that this
model fails to account for merging, and cannot produce
an anti-bias ($b_1 <1$). 
BK99 discuss
a generalization of the Fry model to the case of arbitrary expansion history.
Other, more general models have been
proposed (e.g., Tegmark \& Peebles 1998), but
we employ the generalized Fry 
model as a first approximation to investigate the
effects of bias evolution,
and merely quote the relevant results below.
The bispectrum in this model is given
by
\begin{eqnarray}
     B({k}_{1},{k}_{2},{k}_{3},w) & = & P(k_{1},w)P(k_{2},w)
     \left[ C_1(w)
     + C_2(w) \cos\psi \left(\frac{k_{1}}{k_{2}}+ 
     \frac{k_{2}}{k_{1}}\right) 
     + C_3(w) \cos^2 
     \psi \right] \nonumber \\
     &   & \;\; + \;\; \mbox{(cyc.)},
\label{Bfe}
\end{eqnarray}
where
\begin{eqnarray}
     C_1(w) & = & \frac{(10/7)d^2(w) + 2(b_{1_{*}}-1)[d(w)-2/7] + b_{2_{*}}}
     {[d(w)+b_{1_{*}}-1]^2}, \\
     C_2(w) & = & \frac{d(w)}{d(w)+b_{1_{*}}-1}, \\
     C_3(w) & = & \frac{(4/7)[d^2(w) + b_{1_{*}} -1]}{[d(w)+b_{1_{*}}-1]^2},
\label{Cs}
\end{eqnarray}
$d(w)=D(w)/D(w_{*})$, and a subscripted asterisk denotes
the value of that parameter at the epoch of formation. 
Note that we have now ignored the very weak dependence of the bispectrum
on $\mu$; the dependence on the expansion history, i.e., on
the species contributing to the total energy density, is contained
in the growth factor, $D(w)$.
The result for $s_3$ can then be extended to the case of evolving bias
by using the bispectrum of equation~(\ref{Bfe}) in equation~(\ref{askewness}), 
which amounts to making
the substitutions
\begin{eqnarray}
\frac{1}{b_1} \longrightarrow C_2(w), \:\:\:\:
\frac{(1-\mu)}{b_1} \longrightarrow C_3(w), \:\:\:\:
\frac{b_2}{b_1^2} \longrightarrow C_1(w) -2C_2(w) + C_3(w) 
\end{eqnarray}
inside the brackets in equation~(\ref{askewness2}), and replacing $b_1$
by $1/{C_2(w)}$.

\begin{figure}[htbp]
\plotone{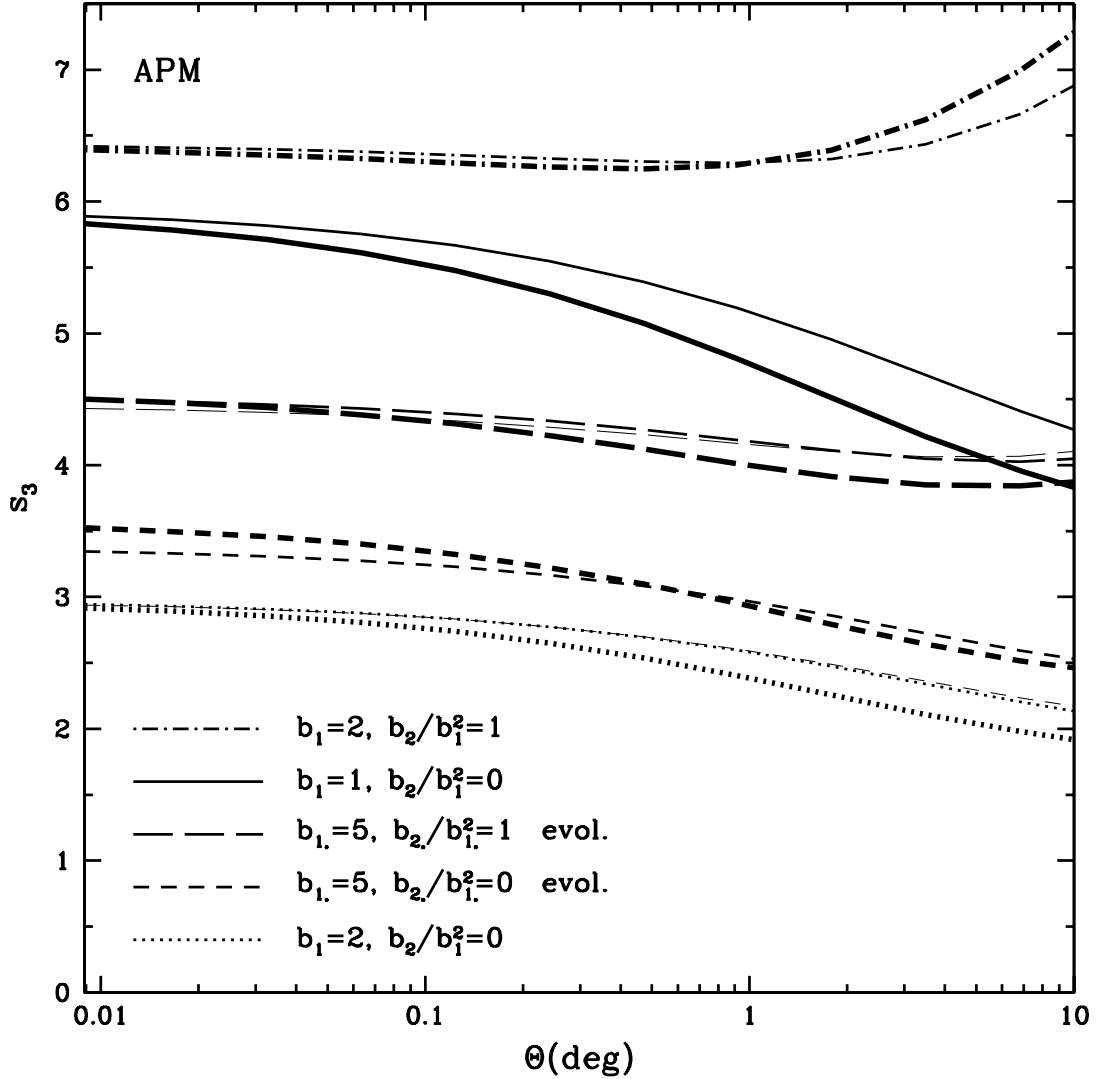}
\caption{Plot of $s_3(\Theta)$, assuming the APM SF, for the five bias 
scenarios described in the text, each employing the SCDM model (thick lines),
$\Lambda$CDM model (medium lines), and OCDM model (thin lines). For the
non-evolving bias cases, the latter two models are taken to be identical,
as suggested by Figure \ref{fig:s3_a12}. Note the strong sensitivity to the
biasing scheme and the presence of the upturn in $s_3$ at large $\Theta$
in the nonlinear bias cases. As expected, linear biasing tends to lower
the predicted curves, while nonlinear biasing tends to raise them.
The evolving bias models yield less dramatic shifts than their
non-evolving counterparts, since evolving bias can effectively act 
as a constant bias, which at late times will be small.}
\label{fig:s3_a12_bias}
\end{figure}
\begin{figure}[htbp]
\plotone{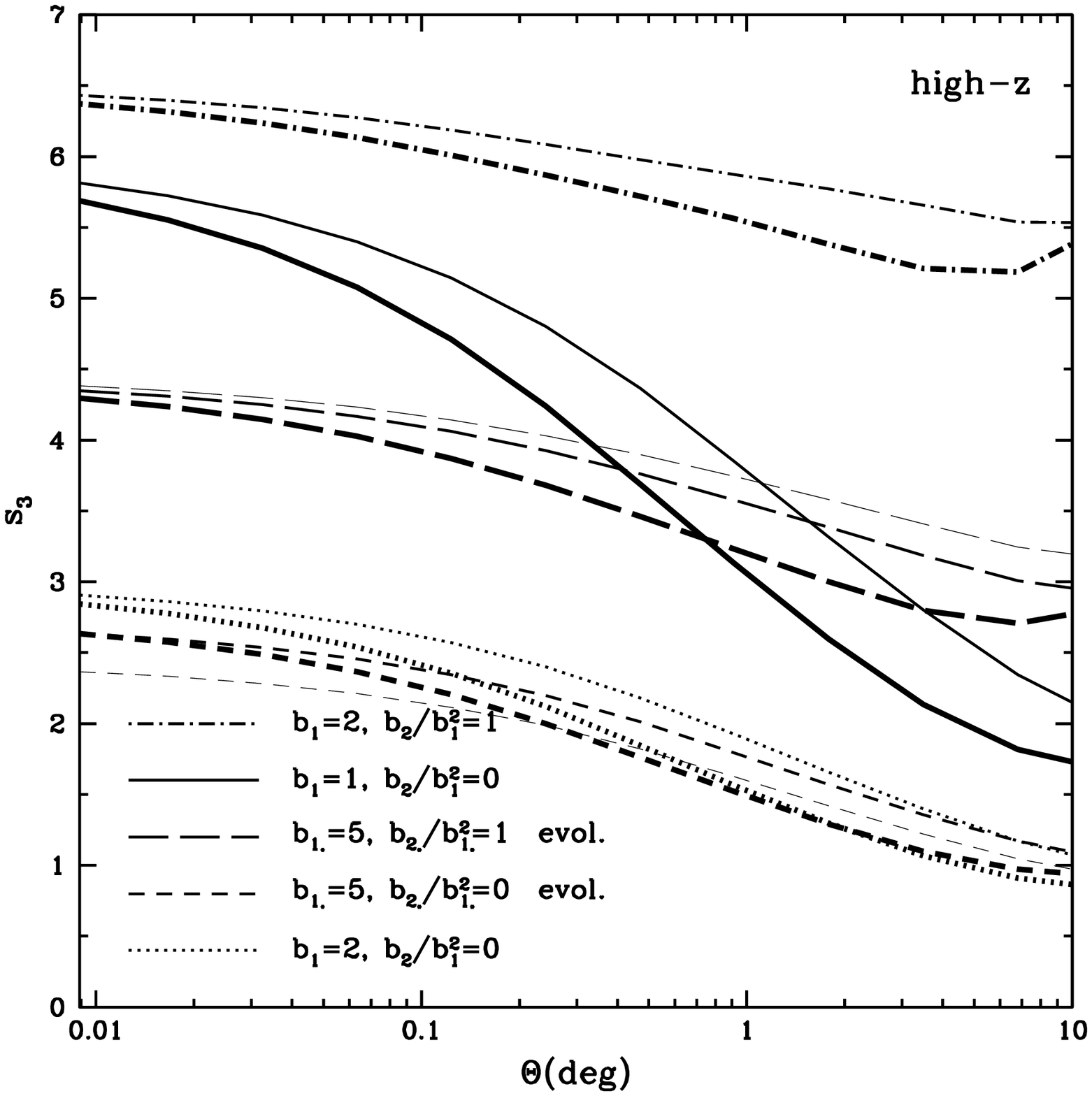}
\caption{Same as Figure \ref{fig:s3_a12_bias}, but using the high-$z$ SF.
Note again the impact of the biasing scheme, the signatures of linear
and nonlinear biasing, and the upturn in $s_3$ seen 
in cases where $b_2 \neq 0$. the Figure also illustrates one example of
the possible degeneracies of $s_3$ with respect to the cosmological parameters,
the bias parameters, and the evolution in the latter. In this case, the 
predictions for the SCDM and $\Lambda$CDM models with 
evolving linear-bias ($b_{1_{*}}=5$; short-dashed
lines) are respectively similar to 
the SCDM and $\Lambda$CDM/OCDM models with
constant, linear bias ($b_1=2$; dotted lines).
}
\label{fig:s3_k_bias}
\end{figure}
For the APM and high-$z$ SFs, respectively,
Figures \ref{fig:s3_a12_bias} and \ref{fig:s3_k_bias} 
display the results for $s_3(\Theta)$ in 
five different bias scenarios, each
employing the SCDM, $\Lambda$CDM, and OCDM models above:  
an unbiased scenario 
($b_1=1$, $b_2/b_1^2=0$, taken from Figures \ref{fig:s3_a12} and 
\ref{fig:s3_k}), 
a non-evolving, linear bias ($b_1=2$, $b_2/b_1^2=0$),
a non-evolving, nonlinear bias ($b_1=2$, $b_2/b_1^2=1$),
an evolving, linear bias ($b_{1_{*}}=5$, $b_{2_{*}}/b_{1_{*}}^2=0$),
and an evolving, nonlinear bias ($b_{1_{*}}=5$, $b_{2_{*}}/b_{1_{*}}^2=1$).
For the non-evolving bias scenarios, the $\Lambda$CDM and OCDM models
are taken to be identical, as suggested by Figures \ref{fig:s3_a12} and 
\ref{fig:s3_k},
and therefore only the $\Lambda$CDM models are plotted in these cases.
These two models do,
however, differ slightly more 
in evolving bias scenarios, due to the different
time dependences of their linear growth factors, and the differences 
between them would increase with decreasing $\Omega_0$.
For the evolving cases we hereafter assume a formation redshift of $z_f = 5$. 

The qualitative dependences of $s_3$ on the bias scenario
are similar to those of its spatial counterpart (BK99). Specifically,
it is clear from these Figures that the dependence of the angular skewness
on the biasing scheme can potentially be far more significant
than that upon the cosmological parameters within a given scheme.
For example, adding a very small linear bias to the low-$\Gamma$
models would yield a result very similar to the unbiased SCDM predictions. 
Generally, the presence of any significant linear
bias,
$b_1 > 1$, tends to reduce the dependence
of $s_3$ on $\Theta$, as compared with the unbiased cases
(solid lines).  Furthermore,
an observed $s_3$ curve which is far below the predicted unbiased result
can only be achieved by a large linear bias term (assuming $b_2 \geq 0$),
while one well above this value can only arise from the existence of a 
non-zero $b_2$ term or from anti-biasing ($b_1 < 1$).
The curves for an evolving bias can produce less drastic shifts
than their non-evolving counterparts, despite the fact that the respective
bias terms are initially larger. 
This is because the evolution towards an unbiased
state can effectively mimic a smaller
(especially at late times), constant bias, as demonstrated by BK99. 

Again we find that the APM curves in Figure \ref{fig:s3_a12_bias} 
are flatter than their 
high-$z$ counterparts in Figure \ref{fig:s3_k_bias}, for reasons 
explained in \S2.1.
Note also that models with nonlinear bias exhibit an upturn at large
$\Theta$.
The upturn arises from the contribution of the higher-order terms
in the sum in equation~(\ref{askewness2}), 
which, in the presence of a non-zero $b_2$ term, can
become significant on large scales, where the effective power-spectrum
index becomes positive (see Figure 1). BK99 perform the analogous three-dimensional
calculation and find similar large-scale behavior of the normalized
spatial skewness, $S_3$.
Depending on the cosmological model, this transition here occurs at angular
scales of about a few degrees for the APM SF, and slightly higher
values in the case of the high-$z$ SF. Thus,
the SCDM model, whose power spectrum
peaks at larger wavenumber than those of
the low-$\Gamma$ models, will exhibit the upturn at smaller
values of $\Theta$, while the low-$\Gamma$
models with 
nonlinear bias are expected to show an upturn 
only at larger angular scales, where they begin to
sample positively-sloped regions of the power spectrum.
However, unlike the case for $S_3$ (BK99), 
the exact location and severity of this upturn in $s_3$
for the different
cosmological models is influenced not only by the shape
of the power spectrum, 
but also by the weightings arising
from the SF (see Figure \ref{fig:dndz}) 
and from the terms proportional to powers of $w$
in equations~(\ref{askewness2}) and (\ref{varpizero}). 
For example, the APM SF, which resembles a low-$w$ spike, exacerbates
the upturn by effectively sampling the
$w$ integrals only around the low peak value of $w(z_m)$, where the 
$w^{-(2n+4)}$ factor is extremely large. In the case of the high-$z$
SF, while large values of $\Theta$ still sample a positively-sloped
region of $P(k)$, the SF itself, which is everywhere much 
shallower than in the APM case, is effectively zero
where the $w^{-(2n+4)}$ factor is large, and peaks at a high value
of $w$, where this factor is very small. As seen in
Figure \ref{fig:s3_k_bias}, this behavior tends to
lend less support to the divergence of
the sum in equation~(\ref{askewness2}) at these scales,
even in the case of a large, constant value of $b_2$.
Note that for both SFs, the non-evolving, nonlinear cases 
exhibit more of an upturn, 
since in these cases the large value of $b_2$ is maintained for
all time, whereas in the evolving cases, the divergence is less severe due
to the fact that the tracer population evolves towards an unbiased
state with time. 

The upturn in $s_3$ at large $\Theta$ in nonlinear-bias models differs
from the decreasing behavior expected if the higher-order terms
in the sum in equation~(\ref{askewness2}) 
are neglected, i.e., if the bias
parameters, as defined for {\em smoothed} density fields in 
equation~(\ref{dexp}), are
are assumed to be constant, as in peak-biasing or halo-biasing models
(Mo \& White 1996). 
BK99 show in the case of the normalized
spatial skewness that this different large-scale behavior might be used to 
obtain better constraints on $b_1$ and $b_2$ on the basis of skewness
measurements alone, since
combinations of linear and nonlinear bias which yield effectively
degenerate results for $S_3$ on smaller smoothing scales, can, 
at least in principle, be distinguished by considering smoothing scales
where the relevant {\em effective} index (Bernardeau 1994) is positive,
i.e., scales greater than $100$ $h^{-1}$ Mpc in typical CDM models
(see Figures 2 and 5 in BK99 and related discussions).
Similarly, we can infer from 
Figures \ref{fig:s3_a12_bias} and \ref{fig:s3_k_bias}
that combinations of these parameters which are 
degenerate
at small $\Theta$ become more distinct at large $\Theta$ (due to the upturn), 
and that comparing QL PT predictions with
large-scale measurements ($\Theta \ga 5^{\circ}$) 
of $s_3$ might also resolve this degeneracy.
If this large-scale behavior is not seen, this would argue in favor
of smoothed-bias prescriptions, rather than the unsmoothed formulation
we employ.
At such scales,
however, one might be limited by the variance induced from the smaller
number of independently sampled cells (Seto \& Yokoyama 1998), 
and may require a more accurate
derivation which does not assume the small-angle approximation, but instead
employs a full spherical harmonic decomposition (Verde \etal 1999).

\begin{figure}[htbp]
\plotone{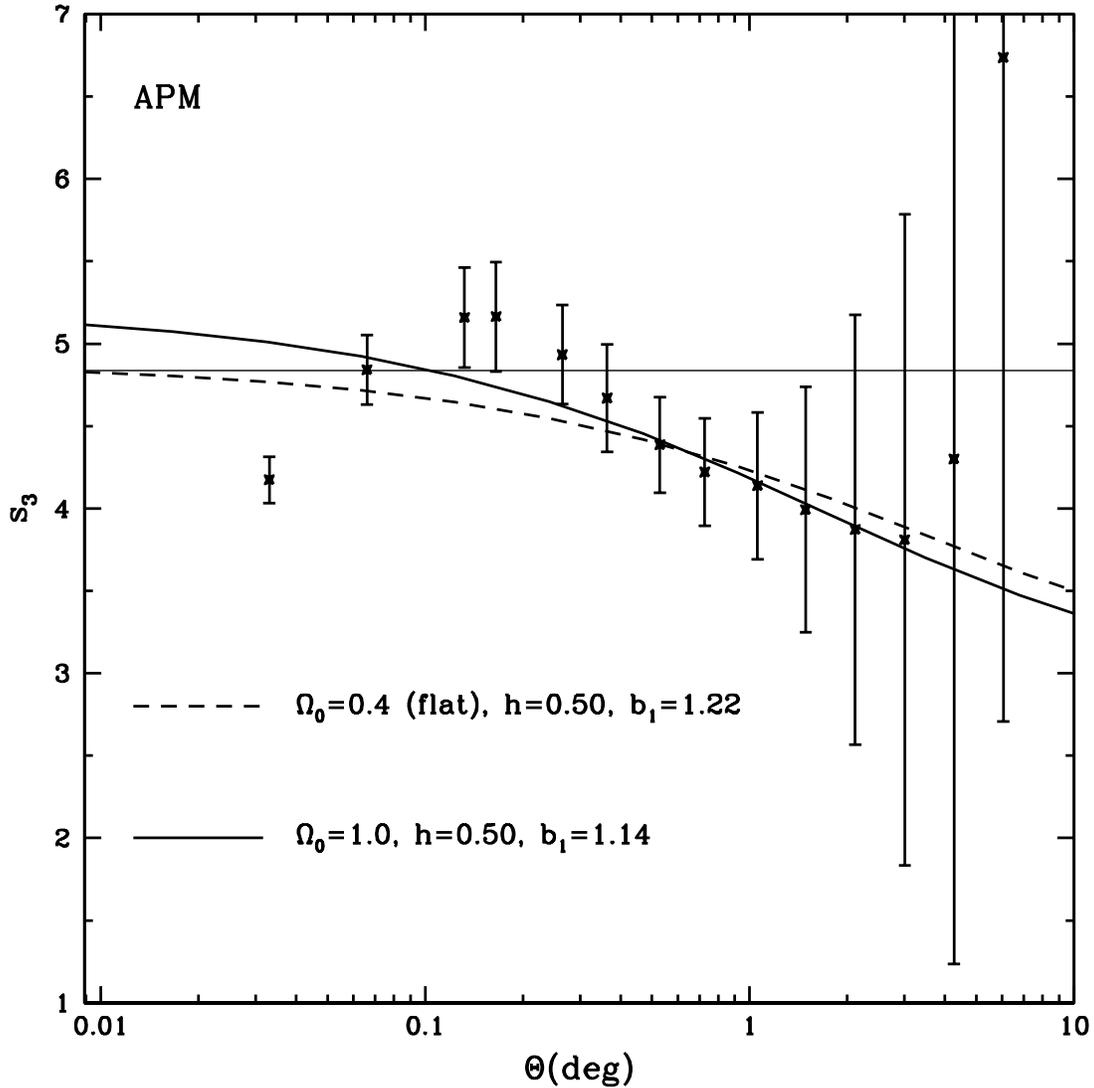}
\caption{Comparison of APM-SF predictions for $s_3$ in the 
SCDM and $\Lambda$CDM
models (with linear bias), with data from the APM Galaxy Survey 
(Gazta\~{n}aga 1994, 1997).
While deviations are expected at small scales (from nonlinear and discreteness
effects) and at large scales (from sample variance and the breakdown of 
the small-angle approximation), the data agree reasonably well, in the QL
regime near $1^{\circ}$, with typical CDM models having a small amount of linear
bias ($b_1 \sim 1.2 \pm 0.1$). The thin horizontal line is taken from
Figure \ref{fig:s3_a12}. }
\label{fig:s3_apm}
\end{figure}

Figures \ref{fig:s3_a12_bias} and \ref{fig:s3_k_bias} also
provide illustrations of one of the potential degeneracies 
that can arise when considering the normalized skewness. 
It is known that the
one-point statistics $S_3$ and $s_3$ can be nearly
degenerate with respect to combinations of 
cosmological and biasing models, with respect to
constant $b_1$ and $b_2$
(barring large-scale measurements, as mentioned above),
and also with respect to evolving vs. non-evolving bias models (BK99).
This latter degeneracy is illustrated in Figure \ref{fig:s3_a12_bias}
by the almost identical predictions of the OCDM model with evolving,
linear bias and the OCDM/$\Lambda$CDM model with constant, linear bias,
and in Figure \ref{fig:s3_k_bias} by the similarity between the
SCDM models with constant, linear bias and with evolving, linear bias.
It would clearly be impossible to distinguish between
such similar predictions given present data of the type shown
in Figure \ref{fig:s3_apm}.

As an example of what can be learned from current measurements, 
we compare in Figure \ref{fig:s3_apm} 
our predictions for $s_3$ using the APM SF
with data from the APM Galaxy
Survey, containing over $1.3 \times 10^6$ galaxies (see Gazta\~{n}aga 1994
and references therein). The
data points (Gazta\~{n}aga 1997) are not corrected for the effects 
of source fragmentation,
and the error bars shown are derived from the scatter in four sub-zones, plus
shot-noise corrections, and are thus extremely conservative, especially at 
large $\Theta$. The curves plotted are for the SCDM and
$\Lambda$CDM models, with ``best-fit'' values for $b_1$ of 1.14 and 1.22,
respectively (assuming only constant, linear biasing). 
It should be noted that discrepancies between the data
and our QL PT predictions are expected at small scales, due
to: 1) shot-noise fluctuations arising from the discreteness of the observed
distribution [i.e., the $\VEV{p_{\Theta}^j}$
are no longer reliable estimators of the area-averaged CFs
(Peebles 1980; Gazta\~{n}aga 1994, 1997)],
2) estimation biases arising from the fact that the ratio of
two unbiased estimators is not itself an unbiased estimator (Hui \& 
Gazta\~{n}aga 1998),
and 3) nonlinear effects not accounted for in our models
(Boucher, Schaeffer, \& Davis 1991; 
Jain \& Bertschinger 1994; Scoccimarro \etal 1998; Scoccimarro 
\& Frieman 1998; Munshi \& Melott 1998).
At large scales we expect discrepancies to arise from
the breakdown of the small-angle approximation
and the large scatter in the data induced by the smaller number of
independently sampled cells. If we therefore exclude the data points
at both the lower and upper ends of the range of $\Theta$, and 
restrict the comparison
to QL scales of $\Theta \la 1^{\circ}$, where our predictions are expected
to be valid, we find reasonable agreement with plausible CDM models
having a small amount of linear bias. Of course, this conclusion is merely
qualitative, as these data are clearly unable to place
tight constraints on the models. 
The 2-$\sigma$ error
on $b_1$ in both models is roughly $\pm 0.1$, depending on exactly which points
are excluded, in agreement with previous determinations
of $b_1 \sim 1.0 \pm 0.2$ for the APM Survey
(Baugh \& Efstathiou 1993; Gazta\~{n}aga 1994; Bernardeau 1995).
That the data for optically selected galaxies are roughly fit in the QL regime
by the predicted values of $s_3$ in CDM models, 
requiring little or no bias (Gazta\~{n}aga 1994; Bernardeau 1995;
Cappi \& Maurogordato 1995; Roche \& Eales 1998; Magliocchetti \etal 1998), 
lends some support to the scenario
of gravitational evolution ensuing from Gaussian initial conditions, 
but better
data are clearly needed.
Note that the thin solid line, derived from equation~(\ref{s3}) with
the APM SF, no bias, and a constant power-law index of $n=-1.2$,
yields a poorer fit in the QL range.

\section{ANGULAR THREE-POINT CORRELATION FUNCTION}

The normalized angular skewness, while
containing some information about the overall
dependence of clustering strength with scale, discards detailed
and valuable 
information about the configuration dependence which is contained in the full
angular 3PCF. It is known that the shape dependence of the spatial
bispectrum and 3PCF, in leading order PT, can be used
to distinguish between the effects of gravitational clustering and bias,
as well as to obtain independent constraints on the linear- and
nonlinear-bias parameters and on cosmological parameters (Jing \& Zhang 1989; 
Fry 1994; Fry 1996; Jing \& B\"{o}rner 1997;
Matarrese, Verde, \& Heavens 1997; BK99). It stands to
reason that the full angular 3PCF should also break these degeneracies
and yield similar constraints.
With this in mind, many authors have investigated the normalized
angular 3PCF,
\begin{equation}
q({\theta}_{12},{\theta}_{23},{\theta}_{31}) = 
\frac{Z({\theta}_{12},{\theta}_{23},{\theta}_{31})}
{\varpi(\theta_{12})\varpi(\theta_{23}) +
 \varpi(\theta_{12})\varpi(\theta_{31}) + 
 \varpi(\theta_{23})\varpi(\theta_{31})},
\label{qdef}
\end{equation} 
where
the angular distances $\theta_{ij}$ form a triangle on the sky.

This quantity, which like $s_3$
is independent of the amplitude of the power spectrum, was introduced
based on empirical arguments for the so-called hierarchical model
(Peebles 1975; Peebles \& Groth 1975; Peebles 1980),
and motivated by data which appeared to indicate that $q$
was simply a constant. 
While equation~(\ref{qdef}) may be taken as a definition,
the result for $q$ in QL PT
is, in general, {\em not} merely a 
constant; we demonstrate that the apparent constancy
of $q$, as measured from past data, 
may simply be an artifact of coarse averaging
over configuration shapes (Peebles \& Groth 1975; Fry \& Seldner 1982), 
combined with the effects of the shallow
SFs which have characterized past surveys. 
There are, however, other problems associated with the definition
of the normalized angular 3PCF in equation~(\ref{qdef}). 
In particular, $q$ is apt to exhibit
rapid variation and divergence where the denominator in equation~(\ref{qdef})
happens to acquire values of or near zero.\footnote{We note that this
problem is avoided entirely if one considers instead the normalized
angular bispectrum in Fourier space.}
This behavior is
artificial in the sense that it does not reflect the true behavior
of the 3PCF itself. BK99 demonstrate the analogous
problems associated with the definition of the normalized spatial 3PCF, $Q$,
and speculate that practical limitations 
in measuring this behavior may, in part,
be responsible for observed discrepancies between QL PT predictions and
those of $N$-body simulations.
The skewness is less susceptible to this problem
since it is scaled by the square of the variance, a positive-definite quantity.
In places where $q$ exhibits this rapidly-varying or divergent behavior,
we will instead consider $q_V$, defined as 
$q_V = Z({\theta}_{12},{\theta}_{23},{\theta}_{31})/(\sigma_{1^{\circ}})^4$,
where $\sigma_{1^{\circ}} \equiv \sqrt{\varpi_{1^{\circ}}(0)}$ is the rms
angular fluctuation on a smoothing scale of $1^{\circ}$, obtained
for each model using equation~(\ref{varpizero}). The chosen smoothing
scale is, of course, arbitrary; we adopt a value of $1^{\circ}$ as a 
convenient choice for the sake of graphical representation.
Like $q$, $q_V$ is
independent of the overall normalization of the power spectrum, but it
is not susceptible to the above mentioned problem associated with
the definition of $q$. 
Observational measurements of both $q$ and $s_3$, however, will necessarily
be ratios of estimators, 
and thus subject to associated errors discussed by Hui \& Gazta\~{n}aga 
(1998).

Taking equation~(\ref{varpi}) with $x = \k\theta$ and omitting the 
window function, since we presume that the full CFs are now measured by
direct counting, rather than by smoothed counts in cells,
we obtain the direct-counting angular 2PCF,
\begin{equation}
\varpi (\theta) = \frac{A}{(2\pi) \theta^{n+2}} 
\int_0^{\eta_{0}} dw \frac{b_1^2 D^2(w)}{w^{n+2}} \left( \frac{dN}{dw} \right)^2
\int_0^\infty dx \: x^{n+1} T^2(x/\theta w) J_0(x).
\label{2pcf}
\end{equation}
The full angular 3PCF for projected triangles with sides of angular measures
$\theta_{12}$, $\theta_{23}$, and $\theta_{31}$
can be evaluated using equation~(\ref{askew}),
\begin{eqnarray}
Z(\theta_{12},\theta_{23},\theta_{31}) & = & \frac{A^2 }{(2\pi)^4}
\int_0^{\eta_{0}} dw \frac{b_1^4 D^4(w)}{w^{2n+4}} \left(\frac{dN}{dw}\right)^3
\int_0^\infty d\k_1 \k_1^{n+1} T^2(\k_1/w) 
\nonumber \\
&\times&  \int_0^\infty d\k_2 \k_2^{n+1} T^2(\k_2/w) 
\int_0^{2\pi} d\phi_1  \int_0^{2\pi} d\phi_2 
e^{i(\bfkappa_1\cdot\bftheta_{13} + \bfkappa_2\cdot\bftheta_{23})} \nonumber \\
&\times& \left[ \frac{1}{b_1} \left( 1+\mu
+ \cos{\psi}\left(\frac{\k_1}{\k_2} +
\frac{\k_2}{\k_1} \right) + (1-\mu)\cos^2{\psi} \right) + \frac{b_2}{b_1^2}
\right] \nonumber \\
&   & \;\; + \;\; \mbox{(cyc.)},
\end{eqnarray} 
where $\phi_i$ is the angle between $\bfkappa_i$ and
$\bftheta_{i3}$, $\psi$ is the angle between $\bfkappa_1$ and 
$\bfkappa_2$, and 
similar conventions are used in each permutation term. 
Taking $x_i = \k_i \theta_{i3}$, and $\phi_{12}$
to be the angle between $\bftheta_{13}$ and $\bftheta_{23}$, so that
\begin{equation}
     \cos{\psi} = \cos{\phi_{12}}(\cos{\phi_{1}}\cos{\phi_{2}} +
     \sin{\phi_{1}}\sin{\phi_{2}}) +
     \sin{\phi_{12}}(\cos{\phi_{1}}\sin{\phi_{2}}
     -\sin{\phi_{1}}\cos{\phi_{2}}),
\end{equation}
and performing the $\phi_1$ and $\phi_2$
integrations, we have
\begin{eqnarray}
Z(\theta_{12},\theta_{23},\theta_{31}) & = & \frac{A^2 }{(2\pi)^2}
\frac{1}{\theta_{31}^{n+2}\theta_{23}^{n+2}}
\int_0^{\eta_{0}} dw \frac{b_1^4 D^4(w)}{w^{2n+4}} \left(\frac{dN}{dw}\right)^3
\int_0^\infty dx_1 x_1^{n+1} T^2(x_1/\theta_{31}w) \nonumber \\
&\times& \int_0^\infty dx_2 x_2^{n+1} T^2(x_2/\theta_{23}w) 
\Biggl\{ \left( \frac{1+\mu}{b_1} + \frac{b_2}{b_1^2} \right)
J_0(x_1)J_0(x_2)  \nonumber \\ 
&  & - \frac{1}{b_1}\cos\phi_{12} \left( \frac{\theta_{23}}{\theta_{13}}
\frac{x_1}{x_2} J_1(x_1) J_1(x_2) + \frac{\theta_{13}}{\theta_{23}}
\frac{x_2}{x_1} J_1(x_1) J_1(x_2) \right)  \nonumber \\ 
&   & + \frac{1-\mu}{b_1} \biggl[ \cos^2{\phi_{12}} J_2(x_1)J_2(x_2)
+\frac{1}{2}\W2(x_1)\W2(x_2) \nonumber \\
&   & - \frac{1}{2}J_2(x_1)\W2(x_2) -
\frac{1}{2}J_2(x_2)\W2(x_1) \biggr] \Biggr\}  \;\; + \;\; \mbox{(cyc.)}.
\label{z}
\end{eqnarray}
For evolving bias we simply substitute
\begin{equation}
\left( \frac{1+\mu}{b_1} + \frac{b_2}{b_1^2} \right)  \longrightarrow  
C_1(w), \:\:\:\:
\frac{1}{b_1}  \longrightarrow  C_2(w), \:\:\:\: 
\frac{1-\mu}{b_1} 
 \longrightarrow  C_3(w)
\end{equation} 
inside the brackets, and replace $b_1$ by $1/{C_2(w)}$.

Equation~(\ref{z}) does not contain terms analogous to the
higher-order terms in the sum in equation~(\ref{askewness2}) 
[which gave rise to the
$\Delta(n)$ term in equation~(\ref{s3})], since evaluation
of the 3PCF by direct counting requires no smoothing. Using our model
for unsmoothed bias, such terms {\em would} be 
induced in the calculation of the 3PCF using
counts in cells, arising from factors of the form $\W2(x\Theta/\theta)$
in the appropriate integrands. One must be careful to realize that
if the unsmoothed-bias prescription is correct, the
nonlinear-bias parameter measured from the direct-counting angular
3PCF in equation~(\ref{z}) will be different from the smoothed
value which could be inferred from $s_3$ or 
from the counts-in-cells 3PCF, though the
two can be related, e.g., via the $\Delta(n)$ term in
equation~(\ref{s3}).

\subsection{Effects of the Power Spectrum}

For a scale-free power spectrum, equation~(\ref{2pcf}) yields
the familiar result, $\varpi(\theta) \propto \theta^{-(n+2)}$.
For realistic CDM power spectra, there
is a further dependence on $\theta$ which arises from the scale-dependence
of the transfer function. Likewise, had we ignored 
the shape-dependent terms in the bispectrum in equation~(\ref{Bf2}), 
we would find for power-law spectra the hierarchical approximation
$Z(\theta_{12},\theta_{23},\theta_{31}) \propto
\varpi(\theta_{12})\varpi(\theta_{23}) +
 \varpi(\theta_{12})\varpi(\theta_{31}) + 
 \varpi(\theta_{23})\varpi(\theta_{31})$. 
These terms, however, introduce into equation~(\ref{z}) an additional, 
non-trivial dependence
on the shape of the triangular configuration, even for the case of
scale-free power spectra. Realistic models must account for this variation,
as well as the scale dependence induced by the transfer function.
\begin{figure}[htbp]
\plotone{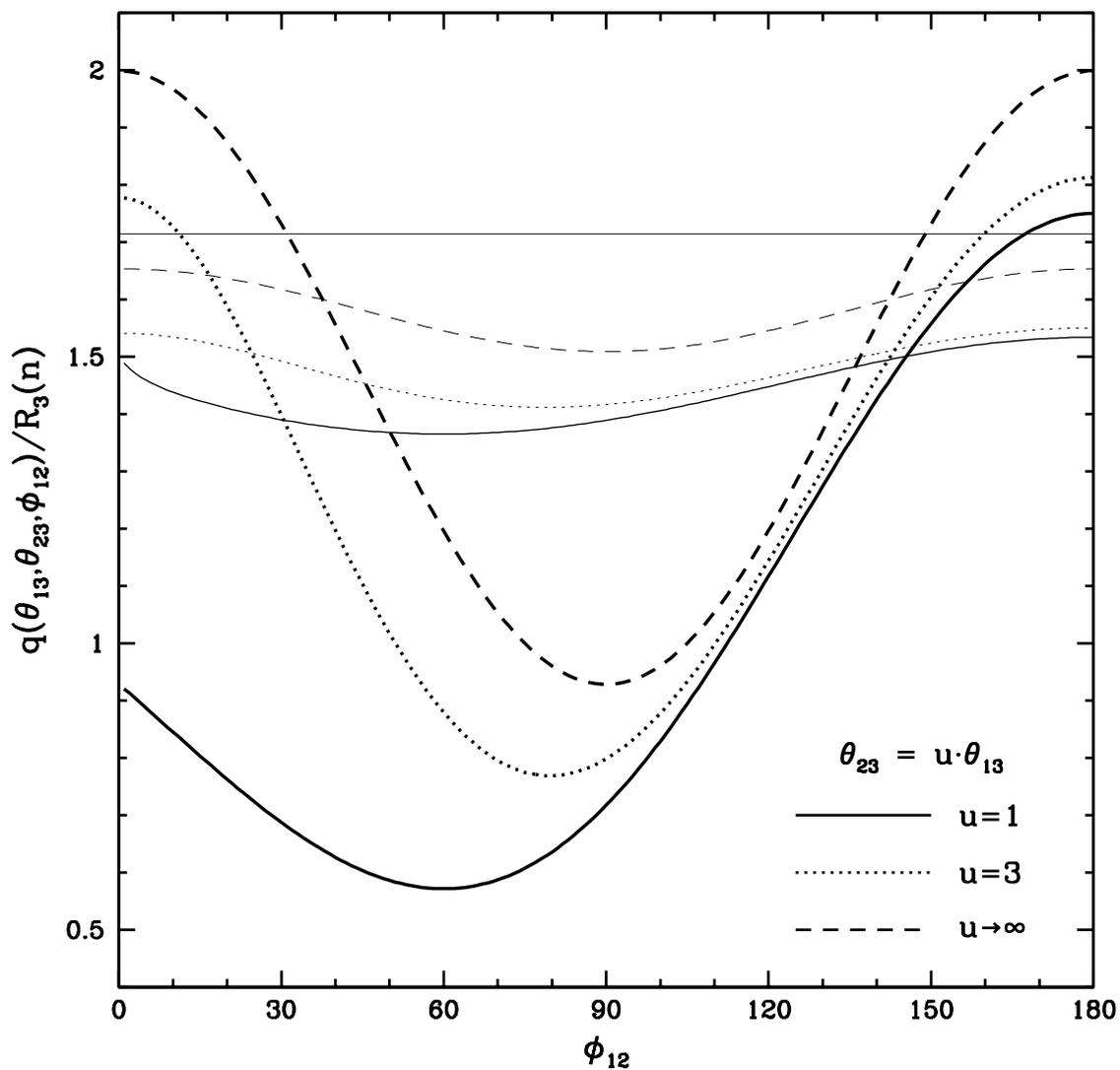}
\caption{Results for $q(\theta_{13},\theta_{23},\phi_{12})/R_3(n)$ 
for power-law spectra
with indices of $n=-1$, (thick lines), $-1.5$ (thin lines), and $-2$ (thin
horizontal line). Since these results are independent of overall scale,
we have fixed $\theta_{13}$ to have length 1, and vary $\theta_{23}$ 
in the manner shown.
Note that the dependence of $q$ on triangle shape varies strongly
with $n$.}
\label{fig:q_powerlaw}
\end{figure}

In order to interpret the results of the various CDM models, it 
will be instructive to first investigate the results obtained for $Z$ for
scale-free power spectra. In this case, we have
\begin{eqnarray}
Z(\theta_{12},\theta_{23},\theta_{31}) & = & R_3(n) \Biggl[ \frac{10}{7}
-\frac{\beta}{2-\beta}\cos\phi_{12}\left(\frac{\theta_{13}}{\theta_{23}} +
\frac{\theta_{23}}{\theta_{13}}\right) \nonumber  \\
&  & + \frac{4}{7}
\frac{(2-2\beta+\beta^2\cos^2\phi_{12})}{(2-\beta)^2} \Biggr]
\varpi(\theta_{13})\varpi(\theta_{23}) + \mbox{(cyc.)},
\label{zpowerlaw}
\end{eqnarray}
where $\varpi(\theta) \propto \theta^{-\beta}$ and $\beta=n+2$.
Figure \ref{fig:q_powerlaw} displays the results for $q/R_3(n)$ for
$n=\{-1, -1.5, -2\}$ [recall $R_3(n)$ is of order unity, equal to
\{1.26, 1.19, 1.15\} and \{1.21, 1.18, 1.17\} for these values of $n$, for
the APM and high-$z$ SFs, respectively]. Since $q$ in this case is
independent of overall scale, we specify the triangle
configuration by fixing $\theta_{13}$ to have length 1,
$\theta_{23}$ to have a relative length $u>1$, and examine the variation of
$q$ with $\phi_{12}$, for $u=1$, $3$, and the ``limit'' 
of $u \rightarrow \infty$. 
The Figure reveals shape dependence on both $u$ and $\phi$, the extent
of which can (particularly in the case of the $\phi$ dependence)
depend strongly on the index $n$. Specifically, we note that the amplitude
of the shape variation tends to increase with increasing $n$.
An ``average'' value for $q$ can be obtained by taking $\VEV{\cos\phi} = 0$
and $\VEV{\cos^2\phi}$ = 1/2 in equation~(\ref{zpowerlaw}), yielding
$\overline{q} = \frac{12}{7}R_3(n)$; while this is not generally valid,
since the three angles appearing in the cyclic terms cannot be varied
independently, this result is exactly true for all configurations when
$\beta=0$ (the no-smoothing case), 
as demonstrated by the thin, solid, horizontal line.

\begin{figure}[htbp]
\plotone{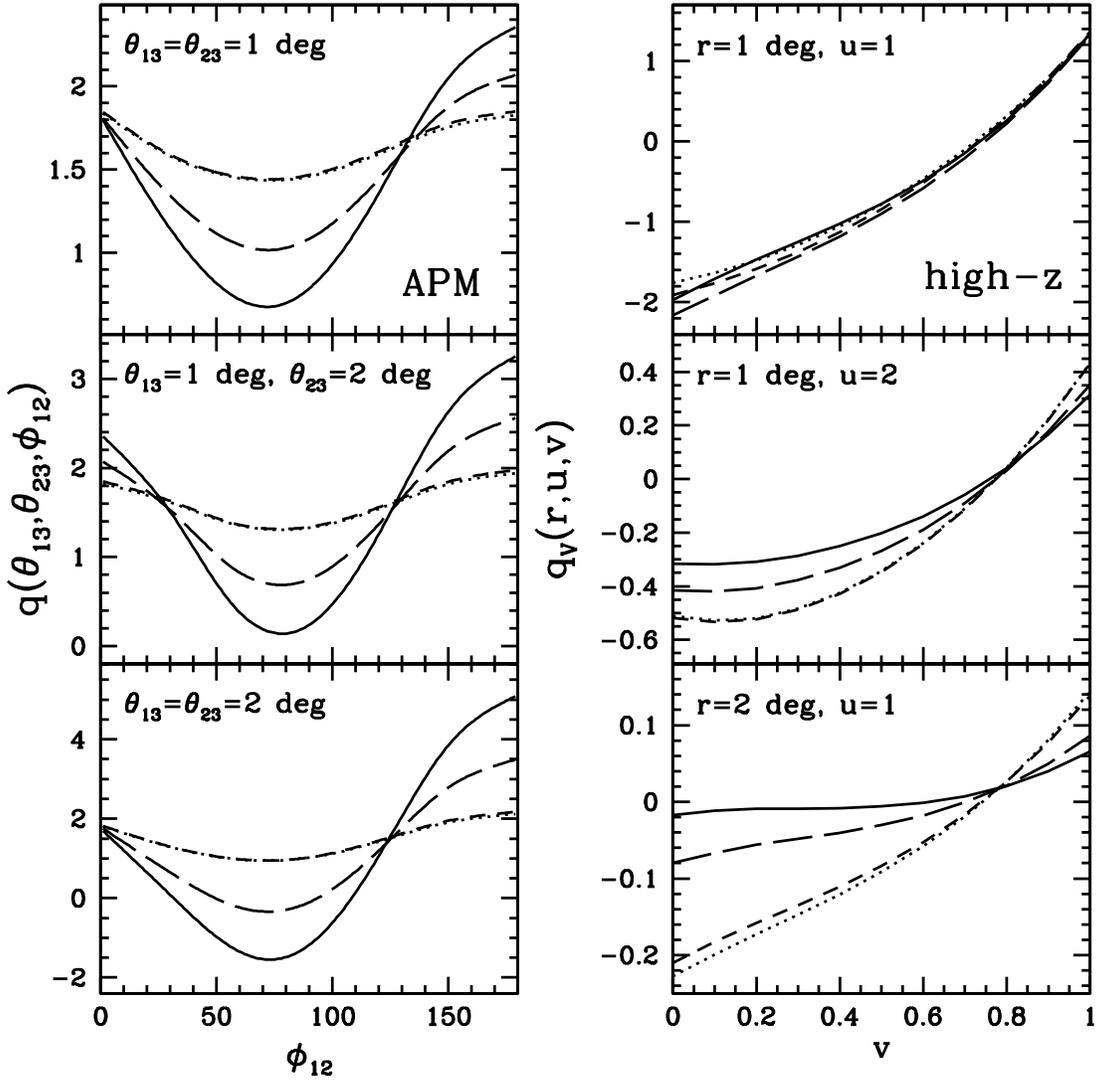}
\caption{The left panels show results for $q$, assuming the APM SF, using the
$\{\theta_{13}, \theta_{23}, \phi_{12}\}$ parameter set to define triangles. 
The right
panels show the results for $q_V$ for the same choice of configurations
(assuming $\phi_{12} > 60^{\circ}$) using the $\{r, u, v\}$ parameter set.
In both cases, the unbiased SCDM (solid lines),
TCDM (long-dashed lines), $\Lambda$CDM (short-dashed lines),
and OCDM (dotted lines) models are
shown. Note that for both SFs, models with different values of $\Gamma$
yield distinct predictions, 
and that the precise configuration dependence of the normalized
3PCF in each different model can vary appreciably with the SF.}
\label{fig:2d}
\end{figure}
In the left panels of Figure \ref{fig:2d} we illustrate,
for the APM SF, the 
behavior of $q(\theta_{13},\theta_{23},\phi_{12})$
for several configurations
falling in the quasilinear regime, for an unbiased tracer mass in
the SCDM, TCDM, $\Lambda$CDM and OCDM models. In this case, the
results for the latter two models differ only by about the thickness of
the curves, and can barely be distinguished. Other models,
however, yield quite distinct predictions and might be
differentiated on the basis
of angular 3PCF measurements from large surveys. 
In particular, the SCDM and TCDM can be clearly distinguished from each other
(BKJ show that this is not the case when considering the spatial 3PCF
at $z=0$), and from the low-$\Gamma$ models, which show 
comparatively little variation
with $\phi_{12}$.
Generally,
we now find a dependence
of $q$ on scale (compare, e.g., the top and bottom panels) arising from the
shape of the CDM transfer function. The observed shape dependence
on $\theta_{23}$ (compare, e.g, the top and middle panels)
and $\phi_{12}$ of these models can be understood in terms
of the results for scale-free power spectra: over the scales
considered here, the SCDM model has the largest effective 
power-spectrum index, while the low-$\Gamma$ models have the smallest.
The results for the APM SF shown
in Figure \ref{fig:2d} roughly agree with the predictions
and $N$-body results of Frieman and Gazta\~{n}aga (1999). We
thank them for pointing out an error in an earlier calculation
of ours.

For a shallow SF such as that describing the APM Galaxy
Survey, three points describing an elongated triangle in projection
are more likely to correspond to a configuration which is elongated
in real space. Thus, our results for $q$ using the APM SF are similar
to corresponding results for the normalized spatial 3PCF, $Q$ (Fry 1984;
Jing \& B\"{o}rner 1997), 
which predict
that clustering in the mildly nonlinear regime favors elongated structures.
For low-$\Gamma$ models, however, the APM SF happens to yield 
predictions which
do not vary strongly with either the shape or size of the
triangular configuration. Since it is known that measurements of the power
spectrum from surveys such as the APM favor models with low values of $\Gamma$
(Efstathiou, Sutherland, \& Maddox 1990;
Baugh \& Efstathiou 1993), it is 
not surprising that measurements of the 
angular
3PCF in shallow surveys have indicated it to be configuration independent,
particularly given the present observational
errors
(Peebles and Groth 1975; Groth \& Peebles 1977; Jing \& Zhang 1989; 
T\'{o}th, Holl\'{o}si, \& Szalay 1989; Jing, Mo, \& B\"{o}rner 1991;
Borgani, Jing, \& Plionis 1991; Jing, \& B\"{o}rner 1998). 
While this has no doubt has lent support
to the empirically-motivated hierarchical model which assumes $Z$ to be, in
fact, shape independent (i.e., $q=$constant), we demonstrate that
the angular 3PCF,
in general, can depend more significantly on the configuration and 
analysis of deeper angular surveys must take
this variation into account.

The results for high-$z$ SF for the same configurations 
(assuming $\phi_{12} > 60^{\circ}$ are shown in the
right panels of Figure \ref{fig:2d}. We now plot $q_V$, since
for this choice of SF the combination of 2PCFs appearing in the 
denominator of $q$ can acquire values of or near zero in our models.
Furthermore, we now take the defining parameters for triangles 
projected on the sky to be
\begin{equation}
r=\theta_{12}, \:\:\:\:\:\: u=\frac{\theta_{23}}{\theta_{12}}, 
\:\:\:\:\:\: v=\frac{\theta_{31}-\theta_{23}}{\theta_{12}},
\end{equation}
where $\theta_{12}<\theta_{23}<\theta_{31}$,
so that $u \ge 1$ and $0 \le v \le 1$ (Peebles \& Groth 1975). 
For this choice
of parameters, $r$ effectively fixes the overall size of the triangle, while
$u$ and $v$ determine the exact size and shape. Figure 6 in 
BK99 provides a graphical
illustration of the variation of triangle geometry with $r$, $u$, and $v$.
This parameterization is convenient since each choice of
\{$r$, $u$, $v$\} involves a minimum scale set by $r$. The \{$\theta_{13}$, 
$\theta_{23}$,
$\phi_{12}$\} parameter set, with $\theta_{13}=r$ and $\theta_{23}=ur$
is equivalent for $\phi_{12}$
ranging down to $60^{\circ}$, but for $\phi$ less than $60^{\circ}$ this
parameter set can sample scales (given by $\theta_{12}$) all the way
down to zero. Since
we have chosen to normalize $Z$ by the square of the variance at a fixed
angular scale, the \{$\theta_{13}$, $\theta_{23}$,
$\phi_{12}$\} set can yield dramatically different amplitudes for
$q_V$ if $0^{\circ} < \phi_{12} < 60^{\circ}$ as opposed to those for
$60^{\circ} < \phi_{12} < 180^{\circ}$, and is therefore less tenable.

Even once their different normalizations have been accounted for,
we find that the high-$z$ curves have significantly lower amplitudes 
than their 
APM-SF counterparts in the left panels. 
This is simply because projected triangles in a deeper
survey can correspond to much larger physical triangles, for
which the clustering amplitude will be significantly smaller. 
Since $q_V$ is normalized by the square of the variance at a fixed
scale, these plots now reveal clearly the expected decrease in clustering
strength with increasing scale; this behavior, however, is not described
simply by the $Z \propto \theta^{-2n+4}$ prediction of the hierarchical model.
Unlike the APM case, we now find that {\em all} models show a significant
variation with shape, as well as scale, with the low-$\Gamma$ models
in some cases now showing the strongest shape dependence.
The SCDM and TCDM models appear less differentiated
than in the APM case, but we find that the 
$\Lambda$CDM and OCDM models can now differ somewhat more substantially,
particularly on larger scales,
since deeper surveys are more sensitive
to cosmological effects.
Most importantly, however, we still find overall that the
high-$\Gamma$ and low-$\Gamma$ models can be fairly well resolved 
on $2^{\circ}$ scales, and the distinction increases for larger scales.
Naturally, a thorough comparison
with measurements spanning the full configuration space in a more
complete fashion would allow for better discrimination
among models.

\subsection{Effects of Bias and its Evolution}

\begin{figure}[htbp]
\plotone{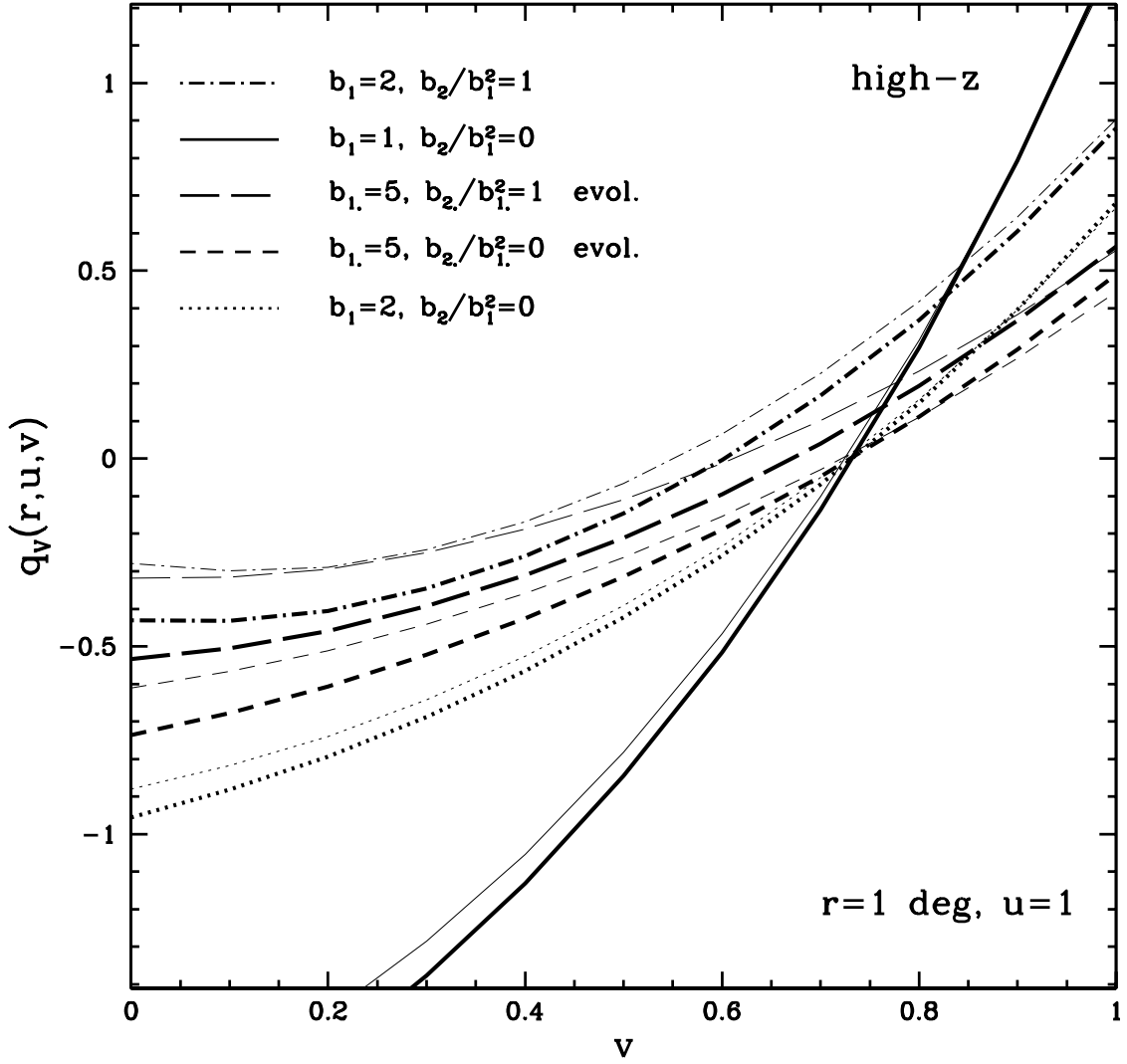}
\caption{Plot of $q_V(v)$, assuming fixed $r=1^{\circ}$ and $u=1$, for the
five bias scenarios from \S2.2, each employing the $\Lambda$CDM (thick lines)
and OCDM (thin lines) models, and using the high-$z$ SF. Note the 
respectively similar influences of the biasing models as were seen 
with $s_3$ (Figures \ref{fig:s3_a12_bias} and \ref{fig:s3_k_bias}).}
\label{fig:z_bias}
\end{figure}
The results for the different bias scenarios investigated in \S2.2
are demonstrated in Figure \ref{fig:z_bias}, which shows $q_V$,
assuming $r=1^{\circ}$ and $u=1$, for the $\Lambda$CDM
(thick lines) and OCDM (thin lines) models, for the high-$z$ SF.
Though this is only one slice of the parameter space
we are exploring, we use it to illustrate the
similar dependences
of the full angular 3PCF on the bias scheme as were seen for $s_3$.
In particular, we see a general flattening and reduction of $q_V$ 
with increasing linear 
bias, and a relative increase with increasing nonlinear bias, as compared
with the corresponding linear scheme.  
Note the slightly different predictions between these two models,
arising from their different expansion histories; these differences 
would be greater
for lower values of $\Omega_0$. As with $s_3$ 
(see Figures \ref{fig:s3_a12_bias}
and \ref{fig:s3_k_bias}), it is clear that measurements
of $q_V$ should be able 
distinguish between the effects of a significant bias
and those arising simply from unbiased gravitational evolution.
 
\begin{figure}[htbp]
\plotone{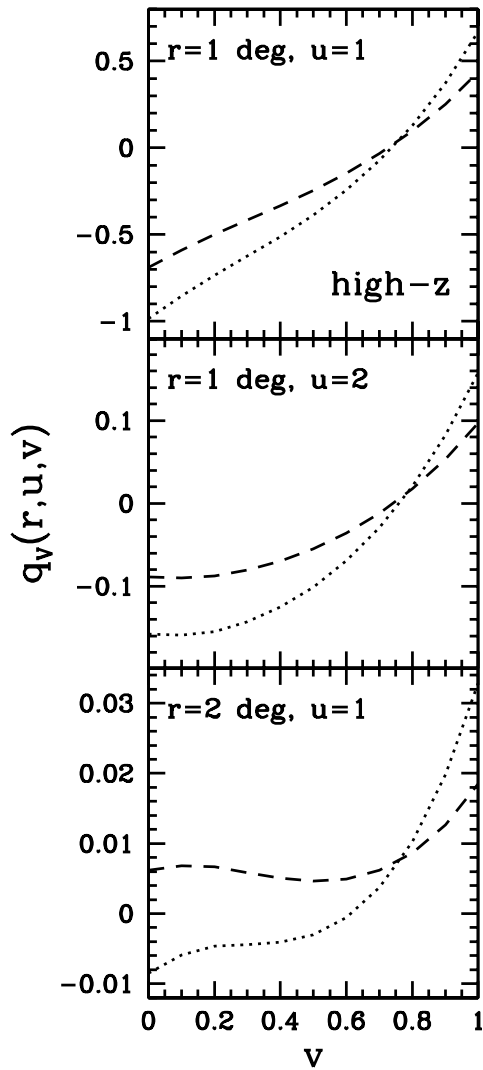}
\caption{Plot of $q_V(r,u,v)$, assuming the high-$z$ SF, for the SCDM
model with constant, linear bias (dotted lines) and the SCDM
model with evolving linear bias
(short-dashed lines) from Figure \ref{fig:s3_k_bias}. While these models
yielded nearly degenerate predictions for $s_3$ in Figure \ref{fig:s3_k_bias},
they are better distinguished here by virtue of the different
configuration dependences of their full angular 3PCFs.}
\label{fig:2dfull_bias_deg}
\end{figure}
Consideration
of the full configuration dependence of the 
angular 3PCF, however, allows for much better discrimination between
different cosmological and biasing schemes
than could be obtained from the skewness alone.
It is known that measurements of the bispectrum and 3PCF
can be used to obtain independent constraints on the constant-valued
linear- and nonlinear-bias parameters 
(Fry \& Gazta\~{n}aga 1993; Fry 1994, 1996; Jing 1997; Matarrese, Verde, \&
Heavens 1997, BK99). This can be inferred here from the different
manners in which $b_1$ and $b_2$ appear in equation~(\ref{z}); the different
weightings of these two parameters is sampled for every triangular 
configuration, producing a graph such as Figure \ref{fig:z_bias} for
every pair of values $(r,u)$ [as opposed to $s_3$, which yields only
one constraint on the combination of $b_1(t)$ and $b_2(t)$].
As another example
of the added information which can be gleaned from the full
geometric dependence of the full 3PCF,
we plot in Figure \ref{fig:2dfull_bias_deg} $q_V(r,u,v)$, 
assuming the high-$z$ SF,
for two of our bias scenarios: the SCDM constant, 
linear-bias model (with 
$b_1=2$ and $b_2=0$), and the SCDM evolving linear-bias model (with
$b_{1_{*}}=5$ and $b_{2_{*}}=0$).
These two cases yielded nearly identical results for $s_3$ in Figure
\ref{fig:s3_k_bias} (where evolving, linear bias effectively approximated
a smaller, constant, linear bias), particularly on scales near $1^{\circ}$.
Here, we see that by considering the variation of $q_V$ with triangle geometry,
the degeneracy between constant and evolving bias
can be alleviated, to a degree, because the angular 3PCF probes
a broad range of redshifts and allows us to see the effects of evolution
in projection. BK99 show
that these evolving and non-evolving bias models might also be 
distinguished
using measurements of $\zeta$ as a function of redshift in a deep
($\overline{z} \ga 1$)
survey, or, to a lesser extent, by
large-scale ($R \ga 20$ Mpc) measurements of $\zeta$ at $z=0$. 
In practice, however, measuring the angular 3PCF
over scales of $\Theta \sim 1^{\circ}$ in
a deep optical survey, or
a radio survey such as the VLA FIRST Survey, may be more feasible and
would likely yield stronger statistical constraints.

The parameter values we have investigated here are only 
intended as illustrations.
The full statistical power which could be gained by
exploring both the configuration and evolutionary
dependences of $Z$ in a 
complete fashion could be useful in
distinguishing between the signatures
of gravitational evolution versus bias, independently measuring the linear-
and nonlinear-bias parameters, and constraining the shape of the power
spectrum. With better data, as might be obtained from emerging
surveys, one could further determine the linear evolution
of the power spectrum and of the bias, which in turn provide
direct constraints on $\Omega_0$, $\Omega_{\Lambda}$, and possibly
the epoch of galaxy formation. However, it is already clear from Figures 
\ref{fig:2d}--\ref{fig:2dfull_bias_deg} that the behavior of
the angular 3PCF is not simply 
that of the hierarchical model which assumes $q$ to be a
shape-independent
constant, and further that, in addition to the full geometric dependence,
one must consider the detailed dependences on the forms and evolutionary
behaviors
of the power spectrum, the biasing mechanism, and the selection function,
as these can have a substantial impact on the results, particularly
for high-redshift surveys.

\section{CONCLUSION}

We have calculated the leading-order results for the normalized angular
skewness and 3PCF, assuming Gaussian initial conditions, 
for an arbitrary, biased tracer-mass distribution in
flat and open universes.
We have considered two selection functions, intended to represent
low-redshift ($\overline{z} \sim 0.1$) and high-redshift
($\overline{z} \sim 1.0$) surveys.
Our derivations incorporate such features
as the scale dependence and linear evolution of the CDM power spectrum 
and the presence of a possibly evolving 
linear or nonlinear bias. Our results for $s_3$ for the case of
an unbiased tracer mass are in agreement with previous work 
(Fry \& Gazta\~{n}aga 1993; Bernardeau 1995) 
and we confirm that PT predictions for $s_3$ in typical CDM models, with
a small amount of linear biasing ($b_1 \sim 1.15 \pm 0.1$), agree
in the QL regime with data from the APM Galaxy Survey. Similar to BK99,
we extend the result for $s_3$ in the case of  
nonlinear bias, to include a
scale-dependent, leading-order correction which becomes significant for
positive effective spectral indices, corresponding to scales
$\Theta \ga 5^{\circ}$ for CDM models. This correction term
can be used to relate the unsmoothed nonlinear-bias parameter
defined in the limit of continuous fields, with the traditionally-used,
scale-dependent, nonlinear-bias parameter defined for smoothed
density fields.
For a given CDM model,
this large-scale behavior of $s_3$ might alone 
allow a more accurate determination
of the linear- and nonlinear-bias parameters
(or at least discriminate between the smoothed and unsmoothed
bias formulations), if other factors, such
as sampling variance and the breakdown of the small-angle approximation,
can be addressed.

We also derive predictions for the normalized angular 3PCF, considering
$q_V$ rather than $q$ in cases where the latter is ill-behaved.
We show that $Z$, as predicted by QL PT, is not given simply
by the form of the ``empirical'' hierarchical model, i.e., with $q$
simply a constant of order unity, but has a more complex
dependence on scale, shape, and expansion history. 
While for shallow SFs the weak
shape dependence of $q$, combined with measurements coarsely averaged
over configuration shapes, can yield results consistent with this 
hierarchical model, ignoring the detailed geometric and evolutionary
variations of the angular 3PCF
in general can destroy much valuable information.
In several cases, we illustrate explicitly
the importance of the various evolutionary
dependences which factor into our calculations of angular statistics, such
as the linear evolution of the power spectrum, the evolution of the bias
parameters, and the $\Omega_0$ dependence of the SFs. The proper treatment
of these detailed dependences will be important in interpreting data from
the emerging generation of deep, high-precision surveys, particularly when
considering the high-redshift universe.

We find that biasing plays a major role in predictions for $s_3$ and $Z$.
In particular, the scale dependence of $s_3$ and the
configuration dependence of $q$ (or $q_V$)
bear characteristic
imprints of bias, such as a relative flattening and decrease with increasing
$b_1$ and a relative increase with increasing $b_2$. 
Compared with the dependence on the biasing scenario, the dependence of
$s_3$ on the cosmological model is fairly weak. This is not the case,
however, for $q$, which can vary appreciably with the values of $\Gamma$ and
$n$.
Though $s_3$ preserves information
about the overall scale dependence of the 3PCF, it provides only one
constraint on the combination of the cosmological parameters,
$b_1(t)$, and $b_2(t)$, leaving various possible
degeneracies between these quantities, including the degeneracy
between constant- and evolving-bias models. 
These can be partially alleviated
by considering the dependence of the normalized angular 3PCF
on the projected-triangle geometry. In particular, the variation
of $q$ can be used to distinguish between the effects
of gravitational evolution and bias, place
constraints on the value of $\Gamma$ and the shape of the power spectrum, 
measure the bias terms $b_1$ and $b_2$ and their time evolution, and perhaps
constrain $\Omega_0$ directly.
Comparing predictions for $s_3$ and $q$
with several data sets 
characterizing different tracer populations could allow
multiple, complementary constraints on these important quantities.

While many of the results obtained here for $s_3$ and $Z$ are
qualitatively similar to the corresponding results obtained by BK99 for
their spatial counterparts, $S_3$ and $\zeta$, the use
of angular statistics offers many practical advantages, such as greater
statistical power and the absence of redshift distortions, which warrant
their separate consideration. Our derivations have also relied
on specific assumptions, such as the chosen bias-evolution model and
the validity of both
the small-angle approximation and the assumption of quasilinearity. 
Based on analogous results from three-dimensional calculations,
we expect that qualitative
features such as the upturn in $s_3$ at large smoothing angles
in the case of nonlinear bias, the reduced configuration (scale)
dependence of $q$ ($s_3$) with increasing linear bias of any form, and
the sensitivity of $q$ to the shape of the power spectrum
will be retained in more general calculations.

The utility of the predictions for $s_3$ and $q$
will of course rely on the precision
with which these statistics can be measured. While current data,
of the type shown in Figure \ref{fig:s3_apm}, is clearly insufficient
to obtain strong constraints on the shape of the power spectrum or
on the bias parameters, it is hoped that future surveys, such as the SDSS,
will greatly improve upon this situation. In particular, if we 
optimistically assume the relevant uncertainties to be predominantly
statistical, we can infer that the errors in measurements of $s_3$ from
the SDSS will be roughly a factor of ten smaller than those in
Figure \ref{fig:s3_apm}.
Of course, the comparison of predictions with two-dimensional 
survey data will require the additional consideration of systematic effects.
Among these are
finite-volume and boundary effects, which can have a significant
impact on degree scales and above (Baugh, 
Gazta\~{n}aga, \& Efstathiou 1995; Szapudi \& Colombi 1996;
Colombi, Szapudi, \& Szalay 1998; Gazta\~{n}aga \& Bernardeau 1998; Munshi
\& Melott 1998), estimation biases (Hui \&
Gazta\~{n}aga 1998; Kerscher 1998), Poisson noise (Peebles 1980; 
Gazta\~{n}aga 1994), 
and sampling variance (Seto \& Yokoyama 1998). 
Gazta\~{n}aga \& Bernardeau (1998) also find that QL PT predictions
for angular statistics may underestimate projection effects,
particularly for APM-type surveys.

In addition, the above calculations can
be extended in several ways. For example, one might consider 
stochastic-biasing models
which may yield different results, particularly on smaller scales 
(Fry \& Gazta\~{n}aga 1993; Catelan \etal 1998; Catelan,
Matarrese, \& Porciani 1998; Taruya, Koyama \& Soda 1998;
Blanton \etal 1998; Col\'{i}n \etal 1998; Taruya \& Soda 1998;
Narayanan, Berlind, \& Weinberg 1998). 
Several authors have also investigated the impact of higher-order
nonlinear terms in PT calculations 
(Jain \& Bertschinger 1994; Scoccimarro \& Frieman 1996a,b;
Scoccimarro \etal 1998; Heavens, Matarrese, \& Verde 1998; Scoccimarro \&
Frieman 1998), also expected to become important on smaller scales,
where $\varpi \ga 1$.
Similar calculations can also
be performed for higher-order angular moments
and $n$-point CFs, such as the kurtosis and four-point correlation function, 
and
so on (Suto \& Matsubara 1994; Bernardeau 1994; Lokas \etal 1995;
Gazta\~{n}aga \& Bernardeau 1998),
which, in principle, provide additional, independent constraints.
The formalism can also be applied to structure-formation
models with non-Gaussian initial conditions, such as topological-defect or
isocurvature models, to investigate how these might 
be distinguished from inflationary models (Luo \& Schramm 1993; Jaffe 1994; 
Fry \& Scherrer 1994; Chodorowski \& Bouchet 1996; White 1998).

\bigskip
We are indebted to J. Frieman and E. Gazta\~{n}aga for helpful
discussions and suggestions, and in particular for their help
in clarifying an earlier problem in the calculation. We also
wish to thank R. Scoccimarro, as well as Licia Verde, 
for numerous helpful comments
and discussions. This work was supported at Columbia by D.O.E. contract
DEFG02-92-ER 40699, NASA NAG5-3091, NSF AST94-19906, and the
Alfred P. Sloan Foundation, and at Berkeley by NAG5-6552.

\appendix
\section{PROJECTED BISPECTRUM}

In this Appendix, we derive a simple Fourier-space analogue of Limber's
equation for the bispectrum of the two-dimensional projection of a
three-dimensional scalar field. Our
derivation is similar to that of Kaiser (1992), 
who calculated the projected (two-point)
power spectrum.  We first consider the case of an Einstein-de Sitter
universe and at the end we generalize the results to open or closed
universes.  Jaffe \& Kamionkowski (1998) discuss a different
application of this formalism to isotropic vector fields.

Consider a three-dimensional field, $\delta(\bfx,w)$, at a
conformal lookback time $w$, and its Fourier transform, ${\tilde
\delta}(\bfk,w)$. We will take $\delta$ to have zero mean (or
subtract the mean explicitly). To project onto the
two-dimensional sky, we apply the ``selection function,''
$q(w)$, and denote the resulting field by 
\begin{equation}
      p(\bftheta) = \int_0^\infty dw\; q(w) \delta(w\theta_1, w\theta_2, w,w),
\end{equation}
where $\bftheta$ is a two-dimensional vector on the sky pointed
at by the three-dimensional unit vector $\hattheta$.  We
consider only small patches, and thus can perform a
two-dimensional Fourier transform on essentially flat areas of
sky, giving ${\tilde p}({\bfkappa})$, where ${\bfkappa}$ denotes
two-dimensional Fourier-space vectors.  
As in Kaiser (1992),  the contribution from a thin shell of width
$\Delta w$ centered at $w_0$ to ${\tilde p}({\bfkappa})$ is
\begin{equation}
  {\widetilde{\Delta p}}({\bfkappa}) = {\Delta w \;q(w_0)\over w_0^2} 
  \int  {dk_3\over2\pi}\; 
  {\tilde \delta}\left({\kappa_1\over w_0},{\kappa_2\over w_0},k_3,w \right)
  j_0\left({k_3\Delta w\over2}\right),  
\label{eq:dpk}
\end{equation}
where the $j_0(x)=\sin{x}/x$ factor arises from the integral of
$\exp({-i {\mbox{\boldmath $k_3 \cdot w$}}})$ over the shell.

We define the three-dimensional bispectrum, the Fourier-space analogue
of the three-point function, by
\begin{equation}
     \VEV{ \tilde \delta(\veck) \tilde \delta(\veck') \tilde
     \delta(\veck'')}= (2\pi)^3 \delta_D(\veck+\veck'+\veck'')
       B(\veck,\veck',\veck'');
\label{eq:bispec3d}
\end{equation}
the two-dimensional analogue is
\begin{equation}
     \VEV{ {\tilde p}(\bfkappa) {\tilde p}(\bfkappa')
     {\tilde p}(\bfkappa'')}= (2\pi)^2 \delta_D(\bfkappa+\bfkappa'+\bfkappa'')
       B_p(\bfkappa,\bfkappa',\bfkappa''),
\label{eq:bispec2d}
\end{equation}
where the dimensionality of the Dirac delta function is obvious
{}from its arguments.

We can form the contribution to the two-dimensional bispectrum
{}from the thin shell,
\begin{eqnarray}
  \VEV{ {\widetilde{\Delta p}}({\bfkappa}) {\widetilde{\Delta
     p}}({\bfkappa}'){\widetilde{\Delta p}}({\bfkappa}'')} &=&
     (2\pi)^2\delta_D({\bfkappa}+{\bfkappa}'+{\bfkappa}'') {\left[q(w_0)\Delta
    w\right]^3\over w_0^4} \int {dk_3\over2\pi}
     {dk_3'\over2\pi}
    \nonumber\\
    &\times&  j_0\left(k_3\Delta w\over2\right)
     j_0\left(k'_3\Delta w\over2\right)
     j_0\left(k''_3\Delta w\over2\right) B(r,r',r''),
\end{eqnarray}
where
\begin{equation}
     r=\sqrt{{\kappa_1^2 + \kappa_2^2 \over w_0^2}+k_3^2}.
\label{eq:pdef}
\end{equation}
Now, we make the same simplifying assumptions as in the original Kaiser
derivation: the shells have width $\Delta w/w\ll 1$ but are thick
compared to wavelengths contributing to $\delta$ 
of interest, so $\Delta
w/w\gg 1/\kappa$. Since the $j_0$ factors have width $\Delta k_3\sim
1/\Delta w$, we can pull the $B$ out of the integral and set
$r=r|_{k_3=0}=\kappa/w_0$; only modes perpendicular to the line of sight
contribute in the small-angle approximation.  Then using the
identity,
\begin{equation}
     \int_{-\infty}^{\infty}\int_{-\infty}^{\infty}\, 
     j_0(x)\, j_0(y) \,j_0(x+y)\,dx \, dy \, = \, \pi^2,
\end{equation}
we can write the contribution from the shell as
\begin{equation}
  \VEV{ {\widetilde{\Delta p}}({\bfkappa}) {\widetilde{\Delta
      p}}({\bfkappa}'){\widetilde{\Delta p}}({\bfkappa}'')}=
  (2\pi)^2\delta_D(
  {\bfkappa}+{\bfkappa}'+{\bfkappa}'') {q(w_0)^3\Delta w\over w_0^4}
  B\left({\kappa\over w},{\kappa'\over w},{\kappa''\over w}\right).
\end{equation}
The bispectrum is a ``cumulant''; that is, contributions to the total
bispectrum simply add. Thus, we can sum up the contributions from
different shells, and convert the sum to an integral:
\begin{equation}
    \VEV {{\widetilde{p}}({\bfkappa}) {\widetilde{p}}({\bfkappa}')
     {\widetilde{p}}({\bfkappa}'') }=
  (2\pi)^2\delta(
  {\bfkappa}+{\bfkappa}'+{\bfkappa}'')\int dw\; {q(w)^3\over w^4}
  B\left({\kappa\over w},{\kappa'\over w},{\kappa''\over w}; w\right);
\end{equation}
with the full definition of the bispectrum, this just gives the desired
formula,
\begin{equation}
\label{eq:bispecformula}
  B_p({\bfkappa},{\bfkappa}',{\bfkappa}'')=\int dw\; {q(w)^3\over w^4}
  B\left({\kappa\over w},{\kappa'\over w},{\kappa''\over w}; w\right)
\end{equation}
(allowing the three-dimensional bispectrum to vary with lookback time,
$w$). Note that the integrand in the formula is weighted by $q^3/w^4$;
this is in contrast to the (two-point) power spectrum version of the
calculation which weights by $q^2/w^2$ [see equation~(\ref{Pp2})]. 
Hence, the bispectrum is
weighted considerably more heavily to late times (small $w$).

To generalize to an open or closed Universe, we must replace the
distances $w$ in the derivation above with the angular-diameter
distance.  Thus in an open or closed Universe one replaces
\begin{equation}
     {1 \over w} \quad \longrightarrow \quad { a_0 H_0
     \sqrt{|1-\Omega_0 -\Omega_\Lambda|} \over 
     S(a_0 H_0 w\sqrt{|1-\Omega_0-\Omega_\Lambda|}) }
\label{eq:replacement}
\end{equation}
in the first three arguments of $B\left(\kappa/ w,\kappa'/
w,\kappa''/w; w\right)$ and in the $w^4$ in the 
denominator in equation~(\ref{eq:bispecformula}), where $
S(x)=\sinh x$ in an open Universe and $S(x)=\sin x$ in a
closed Universe.


\clearpage
 
\begin{references}

\reference{   } Baugh, C. M. \etal 1998, astro-ph/9811222, MNRAS, submitted

\reference{   } Baugh, C. M. \& Efstathiou, G. 1993, MNRAS, 265, 145

\reference{   } Baugh, C. M., Gazta\~{n}aga, E., Efstathiou, 
G. 1995, MNRAS, 274, 1049

\reference{   } Bardeen, J. M. \etal 1986, ApJ, 304, 15

\reference{   } Bartlett, J. G. \etal 1998, Fundamental Parameters in Cosmology,
Les Arcs; Publisher: Editions Frontieres, p. 103

\reference{   } Becker, R. H., White, R. L., \& Helfand, D. J. 1995,
ApJ, 450, 559

\reference{   } Bernardeau, F. 1994, ApJ, 433, 1

\reference{   } Bernardeau, F. 1995, A\&A, 301, 309

\reference{   } Blanton, M. \etal 1998, astro-ph/9807029, ApJ, submitted

\reference{   } Borgani, S., Jing, Y., \& Plionis, M. 1992, ApJ, 395, 339

\reference{   } Bouchet, F. R. \etal 1992, ApJ, 394, L5

\reference{   } Bouchet, F. R. \etal 1995, A\&A, 296, 575

\reference{   } Bouchet, F. R. \& Hernquist, L. 1992, ApJ, 400,25

\reference{   } Bouchet, F. R., Schaeffer, R., \& Davis, M. 1991,
ApJ, 383, 19

\reference{   } Buchalter, A. \& Kamionkowski, M. 1999, ApJ, 521, in press

\reference{   } Cappi, A. \& Maurogordato, S. 1995, ApJ, 438, 507

\reference{   } Catelan, P. \etal 1995, MNRAS, 276, 39

\reference{   } Catelan, P. \etal 1998, MNRAS, 297, 692

\reference{   } Catelan, P., Matarrese, P., \& Porciani, C. 1998,
     ApJ, 502, L1

\reference{   } Chodorowski, M. J. \& Bouchet, F. R. 1996, MNRAS, 279, 557

\reference{   } Col\'{i}n, P. \etal 1998, astro-ph/9809202, ApJ, submitted

\reference{   } Colombi, S., Szapudi, I., \& Szalay, A. 1998, MNRAS, 296, 253

\reference{   } Cress, C. M. \etal 1996, ApJ, 473, 7

\reference{   } Cress, C. M. \& Kamionkowski, M. 1998, MNRAS, 297, 486

%
\reference{   } Efstathiou, G., Sutherland, W. J., \& Maddox, S. J. 1990, Nat,
348, 705

\reference{   } Frieman, J. A. \& Gazta\~{n}aga, E. 1999, in preparation

\reference{   } Frieman, J. A. \& Gazta\~{n}aga, E. 1994, ApJ, 425, 392

\reference{   } Fry, J. N. 1984, ApJ, 279, 499

\reference{   } Fry, J. N. 1994, PRL, 73, 215

\reference{   } Fry, J. N. 1996, ApJ, 461, L65

\reference{   } Fry, J. N. \& Gazta\~{n}aga, E. 1993, ApJ, 413, 447

\reference{   } Fry, J. N., Melott, A. L., \& Shandarin, S. F. 
1993, ApJ, 412, 504

\reference{   } Fry, J. N., Melott, A. L., \& Shandarin,
     S. F. 1995, MNRAS, 274, 745

\reference{   } Fry, J. N. \& Scherrer, R. J. 1994, ApJ, 429, 36

\reference{   } Fry, J. N. \& Seldner, M. 1982, ApJ, 259, 474

\reference{   } Gazta\~{n}aga, E. 1994, MNRAS, 268, 913

\reference{   } Gazta\~{n}aga, E. 1997, private communication

\reference{   } Gazta\~{n}aga, E. \& Baugh, C. M. 1998, MNRAS, 294, 229

\reference{   } Gazta\~{n}aga, E. \& Bernardeau, F. 1998,
     A\&A, 331, 829

\reference{   } Gazta\~{n}aga, E., Croft, R. A. C., \& Dalton, G. B. 1995,
MNRAS, 276, 336

\reference{   } Gazta\~{n}aga, E. \& Frieman, J. A. 1994, ApJ, 437, L13

\reference{   } Goroff, M. H. \etal 1986, ApJ, 311, 6

\reference{   } Gott, J. R., Gao, B. \& Park, C 1991, ApJ, 383, 90

\reference{   } Gradshteyn, I. S. \& Ryzhik, I. M. 1980, Table of Integrals,
Series, and Products, Academic Press, New York

\reference{   } Groth, E. J. \& Peebles, P. J. E. 1977, ApJ, 217, 385

\reference{   } Heavens, A. F., Matarrese, S., \& Verde, L. 1998,
     astro-ph/9808016, MNRAS, in press

\reference{   } Hui, L. \& Gazta\~{n}aga, E. 1998, astro-ph/9810194, ApJ, submitted

\reference{   } Jaffe, A. H. 1994, Phys. Rev. D, 49, 3893

\reference{   } Jaffe, A. H. \& Kamionkowski, M. 1998, 
Phys. Rev. D58 043001

\reference{   } Jain, B. \& Bertschinger, E. 1994, ApJ, 431, 495

\reference{   } Jensen, L. G. \& Szalay, A. S. 1986, 305, L5

\reference{   } Jing, Y. 1997, Proceedings of IAU Symposium 183, 
Cosmological parameters and the evolution of the Universe,
August 18-22, Kyoto

\reference{   } Jing, Y. \& B\"{o}rner, G. 1997, A\&A, 318, 667

\reference{   } Jing, Y. \& B\"{o}rner, G. 1998, ApJ, 503, 37

\reference{   } Jing, Y., Mo, H. J. \& B\"{o}rner, G. 1991, A\&A, 252, 449

\reference{   } Jing, Y. \& Zhang, J. 1989, ApJ, 342, 639

\reference{   } Juszkiewicz, R., Bouchet, F.R., \& Colombi S. 1993, ApJ, 412, L9

\reference{   } Kaiser, N. 1992, ApJ, 388, 272

\reference{   } Kamionkowski, M. \& Buchalter, A. 1998, ApJ, in press

\reference{   } Kerscher, M. 1998, astro-ph/9811300, A\&A, accepted

%
\reference{   } Lokas, E. L. \etal 1995, MNRAS, 274, 730

\reference{   } Loveday, J. \etal 1998, Wide Field Surveys in Cosmology, Paris;
Publisher: Editions Frontieres, p. 317

\reference{   } Luo, X. \& Schramm, D. N. 1993, ApJ, 408, L33

\reference{   } Maddox, S. J., Efstathiou, G., \& Sutherland, W. J. 1996, MNRAS,
283, 1227

\reference{   } Magliocchetti, M. \etal 1998, MNRAS, 300, 257

\reference{   } Magliocchetti, M. \& Maddox, S. J. 1998, astro-ph/9811320, MNRAS, submitted

\reference{   } Martel, H. 1995, ApJ, 445, 537

\reference{   } Matarrese, S., Verde, L., \& Heavens, A. F. 1997,
     MNRAS, 290, 651

\reference{   } Matarrese, S. \etal 1997, MNRAS, 286, 115

%

\reference{   } Mo, H. J. \& White, S. D. M. 1996, MNRAS, 282,347

\reference{   } Munshi, D. \& Melott, A. L. 1998, astro-ph/9801011

\reference{   } Narayanan, V. K., Berlind, A. A., \& Weinberg, D. H. 1998,
astro-ph/9812002, ApJ, submitted

\reference{   } Peacock, J. A. 1997, MNRAS, 284, 885

%
\reference{   } Peebles, P. J. E. 1975, ApJ, 196, 647

\reference{   } Peebles, P. J. E. 1980, The Large-Scale Structure of the 
Universe, Princeton University Press, Princeton

\reference{   } Peebles, P. J. E. \& Groth, E. J. 1975, ApJ, 196, 1

\reference{   } Roche, N \& Eales, S. A. 1998, MNRAS, submitted

%
\reference{   } Scoccimarro, R. 1998, private communication

\reference{   } Scoccimarro, R. \etal 1998, ApJ, 496, 586

\reference{   } Scoccimarro, R., Couchman, H. M. P., 
     \& Frieman, J. A. 1998, astro-ph/9808305, ApJ, accepted

\reference{   } Scoccimarro, R. \& Frieman, J. A. 1996a, ApJ Supp.,
     105, 37

\reference{   } Scoccimarro, R. \& Frieman, J. A. 1996b, ApJ, 473, 620

\reference{   } Scoccimarro, R. \& Frieman, J. A. 1998, astro-ph/9811184, ApJ, accepted

\reference{   } Seto, N. \& Yokoyama, J. 1998, astro-ph/9812221

\reference{   } Steidel, C. \etal 1998, ApJ, 492, 428

\reference{   } Suto, Y. \& Matsubara, T. 1994, ApJ, 420, 504

\reference{   } Szalay, A. S. 1988, ApJ, 333, 21

\reference{   } Szapudi, I. \& Colombi, S. 1996, ApJ, 470, 131

\reference{   } Szapudi, I. \etal 1988, astro-ph/9810190, ApJ, submitted

\reference{   } Taruya, A., Koyama, K., \& Soda, J. 1998, astro-ph/9807005, ApJ, accepted

\reference{   } Taruya, A. \& Soda, J. 1998, astro-ph/9809204, ApJ, submitted

\reference{   } Tegmark, M. \& Peebles, P. J. E. 1998, ApJ, 500, L79

\reference{   } T\'{o}th, G., Holl\'{o}si, J. \& Szalay, A. S. 1989, ApJ, 344, 75

\reference{   } Verde, L. \etal 1999, in preparation

\reference{   } White, M. 1998, astro-ph/9811227, MNRAS, submitted

\reference{   } Watson, G. N. 1966, Theory of Bessel Functions, Cambridge
University Press, Cambridge

\end{references}
\end{document}